\documentclass[aps,twocolumn,pre,showpacs,superscriptaddress]{revtex4-1}%
\usepackage{graphicx}
\usepackage{amssymb}
\usepackage{bm}

\usepackage{epsfig}

\usepackage{comment}
\usepackage{amsmath}
\usepackage{bbm} 
\usepackage{makeidx} 
\usepackage{wrapfig}
\usepackage{float}
\usepackage{pstricks,pst-grad,pst-plot}
\usepackage{pst-node}
\usepackage{graphicx}
\usepackage{dcolumn}
\usepackage{bm}
\usepackage{amsmath}
\usepackage{amssymb}
\usepackage{epsfig}
\usepackage{pst-grad,pst-node,pst-tree,pst-plot}
\usepackage{pstricks}
\usepackage{natbib}
\usepackage{ifthen}

\begin{document}

\title{Effect of disorder on condensation in the lattice gas model on a random graph}

\author{Thomas~P.~Handford}
\author{Alexander~Dear}
\affiliation{Department of Chemistry, University of Cambridge, Cambridge, UK}
\author{Francisco~J.~P{\'e}rez-Reche}
\affiliation{Institute for Complex Systems and Mathematical Biology, SUPA, King's College, University of Aberdeen, Aberdeen, UK }
\author{Sergei~N.~Taraskin}
\affiliation{St. Catharine's College and Department of Chemistry,
University of Cambridge, Cambridge, UK}

\begin{abstract}
The lattice gas model of condensation in a heterogeneous pore system, represented by a random graph of cells, is studied using an exact analytical solution.
A binary mixture of pore cells with different coordination numbers is shown to exhibit two phase transitions as a function of chemical potential  in a certain temperature range.
Heterogeneity in interaction strengths is demonstrated to reduce the critical temperature and, for large enough degree of disorder, divides the cells into ones which are either on average occupied or unoccupied. 
Despite treating the pore space loops in a simplified manner, the random-graph model provides a good description of condensation in porous structures containing loops. This is illustrated by considering capillary condensation in a structural model of mesoporous silica SBA-15.
\end{abstract}

\pacs{68.03.Fg,64.60.My,75.60.Ej,64.60.aq}

\maketitle

\section{Introduction}

Condensation in porous media is an interesting physical phenomenon which is affected by the geometry of the pore space and chemical properties of the solid matrix and contained fluid. 
In particular, the sorption curves for condensed liquid {\it vs} external pressure exhibit hysteresis loops and jerky avalanche-type behaviour~\cite{Lilly1993,Kierlik1998,Gelb1999}.
The shape of the hysteresis loop varies significantly between different materials, and is strongly dependent on the structure of the pores~\cite{Sing1985}.
It is important to know the link between the structure of the pores and the form of the hysteresis loops in order to, e.g. characterise the porous structure of materials from sorption experiments.

There are several theoretical models, including classical continuous theories, local mean-field approaches and lattice-gas based models which predict the shape of the sorption curves based upon the underlying structure~\cite{GreggBOOK1982,Yortsos_1999,Gelb1999}.
In this paper, we use a lattice gas model~\cite{Kierlik1998}, which can be mapped to the random field Ising model~\cite{Imry1975,Nattermann1997,Sethna2001}, so that techniques developed for the latter model can be applied to the study of condensation.
In the lattice gas model, the porous medium is coarse grained into cells which can be in three different states: occupied by matrix, liquid or vapour.
The configuration of matrix cells is quenched, meaning that they do not change state during sorption.
Each of the rest of the cells are allowed to change between the two other states, i.e. they can either be
empty (occupied by vapour) or occupied by liquid.
The states of these cells are characterised by a state variable $\tau$ which takes the values $\tau=0$ or $1$ corresponding to an empty or occupied cell, respectively.

In thermal equilibrium, the cells can change state between the occupied and the empty states, due to thermal fluctuations. 
The relative number of empty and occupied cells depends on the external parameters such as pressure (chemical potential, $\mu$) and temperature, $T$.
When the cell is occupied by liquid, the liquid interacts with the matrix.
The strength of this interaction, $w^{\text{mf}}$, is a parameter of the lattice gas model.
Two neighbouring cells, both occupied by liquid, can also interact, and the strength of this interaction, $w^{\text{ff}}$, is another parameter of the model.

Real porous materials are typically heterogeneous in structure and chemical composition~\cite{Wallacher2004,Feng2008,Maddox1997,Edler1998,Fenelonov2001,Sonwane2005}.
In order to account for this heterogeneity we allow the parameters of the lattice gas model to vary between cells.
In addition, the topology of a porous network can be far from a regular lattice~\cite{Schaefer1986,Monson2012,PerezRechePRL2012}, and below it is modelled by a random graph with a fixed degree distribution~\cite{NewmanBOOK,Dorogovtsev2008}.
The main aim of this paper is to take into account the disorder in the interaction strengths and topology, and determine the effects of such disorder on sorption.
In order to do this, we analyse the exact equilibrium solution to the model in the limit of infinite system size, and perform Metropolis dynamics simulations to confirm the analytical findings.
Our analytical technique is based on known recursive techniques for the random-field Ising model on a Bethe lattice~\cite{BruinsmaPRB1984,Mezard2001,Sokolovskii2003}, which we extend to introduce disorder in all three parameters, $w^{\text{mf}}$, $w^{\text{ff}}$ and node coordination number (degree), $q$. 
At the same time, the exact results presented here for condensation are relevant to the random-field Ising model.

Our main findings are the following.
We find an exact analytical solution in equilibrium for the distribution of fluid density, incorporating all three types of disorder, in the form of an integral equation.
This integral equation can be solved numerically, and its solution has been supported using Metropolis dynamics simulations.
We first show that the solution of the model reproduces the expected behaviour for chemically and topologically homogeneous porous media with fixed value of $w^{\text{mf}}$, $w^{\text{ff}}$ and $q$ for all cells. 
As the chemical potential is varied, the system undergoes an equilibrium discontinuous phase transition between liquid and gas phases provided temperature is smaller than a critical value, $T_{\text{C}}$, which depends on the coordination number of the cells. 
Above the critical temperature, there are strong thermal fluctuations and the mean density is a continuous function of $\mu$.
We then analyse separately the consequences of several types of disorder on sorption: in coordination number, matrix-fluid interaction and fluid-fluid interaction.
Within the equilibrium approach, disorder in coordination number is shown to lead to the appearance of several first-order phase transitions in a certain range of temperatures as the chemical potential is varied, each transition being associated with a sudden change in the occupation of cells of a certain coordination number.
Dynamically, the system follows a hysteresis curve involving a sequence of metastable states.
This hysteresis loop can have an asymmetric shape, with e.g. two steps for desorption curve and a single step for the adsorption curve.
Disorder in the interaction parameters reduces the critical temperature of the transitions.
For large disorder, the disorder-controlled distribution of the mean occupations of cells is shown to be bimodal, corresponding to either empty or occupied cells.

The existence of more than one phase transition in Ising-like models has previously been observed as a result of disorder in either coordination number~\cite{Vakarin2007} or bimodal disorder in interaction strengths~\cite{Andelman1983,Swift1994}.
In addition, evidence of multiple step transitions in sorption has been found in molecular dynamics simulations~\cite{Page1996}.
However, only one phase transition has been reported in a tree-like system with random cells blocked by matrix~\cite{Sokolovskii2003}, equivalent to a binomial distribution of coordination numbers.
A more general distribution of coordination numbers allows a second phase transition to occur in a tree-like system, as we show below.

Finally, we have demonstrated that the random-graph model can be successfully used for description of sorption in models of realistic porous materials such as SBA-15 when the temperature exceeds the critical one. 
This is due to the fact that, in this temperature regime, the size of avalanches in fluid density is smaller than the typical size of the loops in the pore space and thus the loops are insignificant for condensation.

The structure of the paper is the following.
The model is introduced in Sec.~\ref{sec:Model} and solved in Sec.~\ref{sec:Solution}. 
The results are presented in Sec.~\ref{sec:zero-disorder} for the system without disorder and in Sec.~\ref{sec:CoordDis} for disorder only in coordination numbers. Results for systems with homogeneous coordination number but disordered interaction strengths are presented in Sec.~\ref{sec:DisorderInteractions}. 
In Sec.~\ref{sec:AllDisorder} results for systems with all kinds of disorder simultaneously are presented.
The Metropolis algorithm used to support the analytical findings is discussed in Sec.~\ref{sec:Metropolis}. 
The applicability of the random-graph model for description of sorption in a structural model of real porous materials, such as SBA-15, is analysed in Sec.~\ref{sec:SBA-15} and 
conclusions are presented in Sec.~\ref{sec:Conclusions}.

\section{Model}\label{sec:Model}

Condensation in a porous material with a fixed matrix morphology can be modelled with the following lattice-gas Hamiltonian~\cite{Kierlik1998},
\begin{equation}
{\cal H}=-\sum_{\langle ij\rangle}w^{\text{ff}}_{ij}\tau_i\tau_j-\sum_iw^{\text{mf}}_i\tau_i-\mu\sum_i\tau_i~,\label{eq:Hamiltonian}
\end{equation} 
where $\tau_i$ is the state variable of cell $i$, $w^{\text{ff}}_{ij}$ stands for the strength of fluid-fluid interaction between neighbouring cells $i$ and $j$ and $w^{\text{mf}}_i$ is the strength of fluid-matrix interaction of liquid in a cell with surrounding matrix. 
The network of cells representing the porous material is chosen to be a random-graph of $N$ nodes $i=1,\ldots,N$ with a fixed degree distribution, i.e. there is a fixed number, $N_q$, of nodes of coordination number $q$~\cite{NewmanBOOK}.
The distribution of $w^{\text{mf}}_i$ is assumed to be dependent on the coordination number of the cell $i$, i.e. it is characterised by the probability distribution function (p.d.f.), $W^{\text{mf}}_{q_i}(w^{\text{mf}}_i)$. 
Conversely, the fluid-fluid interaction is assumed to be independent of both $q_i$ and $w^{\text{mf}}_i$ and is distributed according to the p.d.f. $W^{\text{ff}}(w^{\text{ff}}_{ij})$ for all pairs of neighbouring cells.

The relevant characteristics of the system are the mean free energy per cell (or mean free energy per cell using the language of the random-field Ising model), $F/N$, and the mean occupied volume, $\rho_i=\langle\tau_i\rangle$, of each cell, where $\langle\ldots\rangle$ represents an average over the grand canonical ensemble.
These quantities can be defined in terms of the (grand) partition function (with $\beta=T^{-1}$ being the inverse temperature), 
\begin{equation}
Z=\sum_{\{\tau_i\}}\exp(-\beta{\cal H}(\{\tau_i\}))~.\label{eq:Partition}
\end{equation}
according to,
\begin{equation}
F=-\beta^{-1}\ln(Z)~,\label{eq:FreeEnergyDef}
\end{equation}
and,
\begin{equation}
\rho_i=\frac{\partial F}{\partial w^{\text{mf}}_i}~.
\end{equation} 
As a result of the quenched disorder, the values of $F$ and $\rho_i$ are random quantities, with the free energy taking a mean value per cell ${\overline F}/N$ and the mean occupied volume of each cell having a distribution $R(\rho_i)$.
Our aim is to calculate the p.d.f. $R(\rho_i)$ and ${\overline F}/N$ and thus determine how the mean density,
\begin{equation}
{\overline \rho}=\int_{\rho_i=0}^1\rho_iR(\rho_i)\text{d}\rho_i~,
\label{eq:Mean-rho_def}
\end{equation}
depends on $\mu$.

\section{Solution}\label{sec:Solution}

\begin{figure} 
\includegraphics[width=8.6cm,clip=true]{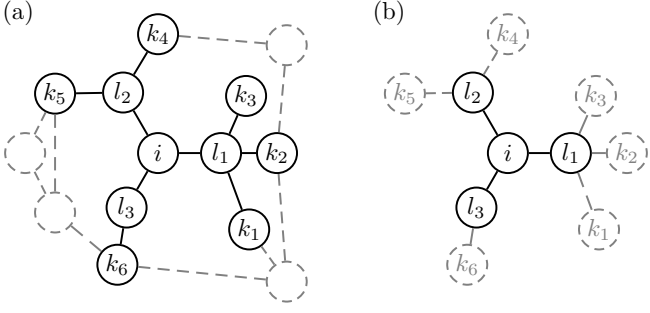}
\caption{Panel (a): Random graph of fixed degree distribution with several loops. 
The cells drawn with dashed circles are assumed to be a long enough distance from cell $i$ that they do not significantly affect it and can be removed. 
Panel (b): The same random graph with distant cells removed to make it a tree. 
The grey dashed cells marked $k_1,k_2,\ldots$ are on the edge of the tree 
and their effect on the cells marked $l_1,l_2,l_3$ neighbouring cell $i$ can be accounted in terms of effective fields, $w^{\text{eff}}_{l_1k_1},w^{\text{eff}}_{l_1k_2},\ldots$, given by Eq.~\eqref{eq:EffFieldEdge}.\label{fig:Tree} 
}
\end{figure}

In this section, we present a method for the calculation of the free energy per cell, ${\overline F}/N$, and the distribution of mean densities, $R(\rho_i)$, on a random graph of fixed degree distribution. 
Such a graph is locally tree-like~\cite{Marinari2004} meaning that the analysis can be performed using the same techniques as applied to a Bethe lattice.

In this analysis, we consider an arbitrary cell $i$, and calculate the distribution $R(\rho_i)$ for that cell and the value of the mean contribution of that cell to the free energy.
The first step in this analysis is to transform the system into a tree by removing
all the cells in the system which are too remote to have a significant effect on the state of cell $i$ (see grey dashed cells in Fig.~\ref{fig:Tree}(a)).
After this step, there will be many boundary cells, $\{k\}$, which are at the edge of the tree and linked only to a single other cell, $l$ (see grey dashed cells labelled $k_1, k_2, \ldots$ in Fig.~\ref{fig:Tree}(b)). 
Following the standard procedure described in~\cite{Katsura1979,Nakanishi1981,Mezard2001,Sokolovskii2003}, the sum over the states $\tau_k=0,1$ of a boundary cell, $k$, in Eqs.~\eqref{eq:Partition} and~\eqref{eq:FreeEnergyDef} for the free energy can be explicitly performed,
\begin{widetext}
\begin{eqnarray}
\beta F&=&-\ln\left[\sum_{\{\tau_i:i\neq k\}}\sum_{\tau_k=0,1}\exp\left(\beta\sum_{\langle ij\rangle}w^{\text{ff}}_{ij}\tau_i\tau_j+\beta\sum_iw^{\text{mf}}_i\tau_i+\beta\mu\sum_i\tau_i\right)\right]\nonumber\\
&=&-F^+\left(-\beta\mu-\beta w^{\text{mf}}_k 
\right)-\ln\left[\sum_{\{\tau_{i}:i\neq k\}}\exp\left(\beta\sum_{\langle ij\rangle:i,j\neq k}w^{\text{ff}}_{ij}\tau_i\tau_j+\beta\sum_{i\neq k}w^{\text{mf}}_i\tau_i+\beta\mu\sum_{i\neq k}\tau_i+\beta w^{\text{eff}}_{lk}\tau_l\right)\right]~,\label{eq:SumPartition}
\end{eqnarray}
\end{widetext}
where the effective field, $w_{lk}^{\text{eff}}$, at cell $l$, due to cell $k$, obeys the following relation,
\begin{equation}
\beta w_{lk}^{\text{eff}}=F^+\left(F^+\left(\beta\mu+\beta w^{\text{mf}}_k\right)+F^-\left(\beta w_{kl}^{\text{ff}}\right)\right)~,\label{eq:EffFieldEdge}
\end{equation}
and the functions $F^{\pm}(x)$ are defined as,
\begin{equation}
F^{\pm}(x)=\pm\ln(\exp(\mp x)\pm 1)~.
\end{equation}
In Eq.~\eqref{eq:SumPartition}, the sum over $\{\tau_i:i\neq k\}$ is taken over all possible states of the vector $\{\tau_1,\ldots,\tau_{k-1},\tau_{k+1},\ldots,\tau_N\}$ representing the states of all cells except cell $k$, while $\langle ij\rangle:i,j\neq k$ means the sum over all neighbouring pairs $i$ and $j$ excluding pairs which include cell $k$.

A similar summation to that performed in Eq.~\eqref{eq:SumPartition} to remove cell $k$ can be done over the states of all the other neighbours of cell $l$ except the cell $m$ which is closer to cell $i$ than cell $l$.
After this, the sum over the state, $\tau_l$, of cell $l$ can be performed, removing cell $l$ and giving an effective field at cell $m$,
\begin{equation}
\beta w_{ml}^{\text{eff}}=F^+\left(F^+\left(\beta w^{\text{part}}_{lm}\right)+F^-\left(\beta w_{ml}^{\text{ff}}\right)\right)~,\label{eq:EffField}
\end{equation}
where the partial local effective field $w^{\text{part}}_{lm}$ at cell $l$, excluding interactions with cell $m$, is given by,
\begin{equation}
w^{\text{part}}_{lm}=\mu+w^{\text{mf}}_l+\sum_{k/l,k\neq m}w^{\text{eff}}_{lk}~.\label{eq:PartField}
\end{equation}
In this expression, the summation 
$k/l,k\neq m$ is over 
all cells $k$ which are neighbours of $l$, not including cell $m$.
The relation given by Eqs.~\eqref{eq:EffField} and~\eqref{eq:PartField} can be used to determine an effective field, $w_{ml}^{\text{eff}}$, acting on cell $m$ and accounting for the effect of some neighbouring cell $l$, closer to the edge of the tree.
Since these relations give $w_{ml}^{\text{eff}}$ in terms of similar effective fields, $w_{lk}^{\text{eff}}$, at cells further out, they can be used to recursively replace all cells except cell $i$ with a single effective field for each neighbour of cell $i$.

Using the effective fields calculated above, the one-cell and two-cell free energies~\cite{Katsura1979} can be defined as,
\begin{eqnarray}
\beta F^{(1)}_{i}&=&-\ln\left[\sum_{\tau_i=0,1}\exp\left(-\beta{\cal H}_i^{(1)}(\tau_i)\right)\right]~,\label{eq:FreeEnergyOneSite}\\
\beta F^{(2)}_{ij}&=&-\ln\left[\sum_{\tau_i,\tau_j=0,1}\exp\left(-\beta{\cal H}_{ij}^{(2)}(\tau_i,\tau_j)\right)\right]~.\label{eq:FreeEnergyTwoSite}
\end{eqnarray}
in terms of the effective Hamiltonians for one site,
\begin{eqnarray}
{\cal H}_i^{(1)}(\tau_i)&=&-\left(w^{\text{mf}}_i+\sum_{j/i}w^{\text{eff}}_{ij}+\mu\right)\tau_i~,\label{eq:HamiltonianOneSite}
\end{eqnarray}
and for two sites,
\begin{eqnarray}
&&{\cal H}_{ij}^{(2)}(\tau_i,\tau_j)=-\left(w^{\text{mf}}_i+\sum_{k/i,k\neq j}w^{\text{eff}}_{ik}+\mu\right)\tau_i\nonumber\\
&&-\left(w^{\text{mf}}_j+\sum_{k/j,k\neq i}w^{\text{eff}}_{jk}+\mu\right)\tau_j-w^{\text{ff}}_{ij}\tau_i\tau_j~.\label{eq:HamiltonianTwoSite}
\end{eqnarray}
In Eq.~\eqref{eq:HamiltonianOneSite}, the sum over $j/i$ represents the sum over all neighbours $j$ of cell $i$.
Following known methods~\cite{Katsura1979,Nakanishi1981,Mezard2001}, the total free energy is then given by equating 
two 
expressions for the Hamiltonian, 
\begin{eqnarray}
\frac{\partial}{\partial\beta}\sum_{i}(q_i-1)\beta F^{(1)}_i-\frac{\partial}{\partial\beta}\sum_{\langle ij\rangle}\beta F^{(2)}_{ij}
&=&\langle{\cal H}\rangle\nonumber\\
\frac{\partial (\beta F)}{\partial\beta}&=&\langle{\cal H}\rangle\nonumber
\end{eqnarray}
implying that,
\begin{equation}
F=\sum_{i}(q_i-1) F^{(1)}_i-\sum_{\langle ij\rangle} F^{(2)}_{ij}~,\label{eq:FreeEnergy}
\end{equation}
and similarly, the mean density at a cell $i$ is given by,
\begin{equation}
\rho_i=\frac{\partial {F}^{(1)}_i}{\partial w^{\text{mf}}_i}~.\label{eq:Rho}
\end{equation}
Eqs.~\eqref{eq:FreeEnergy} and~\eqref{eq:Rho} can be used to calculate $F$ and $\rho_i$ in terms of the effective fields $w^{\text{eff}}_{ij}$ defined above.

The effective fields $w^{\text{eff}}_{ij}$ for each pair of cells $i$ and $j$ are, in general, different, due to the disorder in coordination number and disorder in interaction strengths $w^{\text{ff}}_{ij}$ and $w^{\text{mf}}_i$.
It is known for disordered ferromagnetic systems~\cite{Mezard2001} that the effective fields $w^{\text{eff}}_{ij}$ at any cell $i$ deep inside the tree due to each neighbour $j$ closer to the edge of the tree are identically and independently distributed according to the same p.d.f. $W^{\text{eff}}(w^{\text{eff}}_{ij})$.
Using Eq.~\eqref{eq:EffField}, this p.d.f. is be found to obey the following equation,
\begin{eqnarray}
&&W^{\text{eff}}(w^{\text{eff}}_{ij})\nonumber\\&&=
\sum_{q_j}{\tilde p}_{q_j}\int
\delta\left(\beta w^{\text{eff}}_{ij}-F^+\left(F^+\left(\beta w^{\text{part}}_{ji}\right)\right.\left.+F^-\left(\beta w_{ij}^{\text{ff}}\right)\right)\right)
\nonumber\\&&\times \left[W^{\text{part}}_{q_j}(w^{\text{part}}_{ji})\text{d}w^{\text{part}}_{ji}\right]\left[W^{\text{ff}}(w^{\text{ff}}_{ij})\text{d}w^{\text{ff}}_{ij}\right]~,
\label{eq:RecursiveEquation}
\end{eqnarray}
Here, $q_j$ is the coordination number of the neighbour $j$ of cell $i$, which, for a random graph of fixed degree distribution, takes a value $q_j=q$ with probability ${\tilde p}_{q}=qN_{q}/\sum_{q=1}^{q_{\text{max}}}q{N_q}$~\cite{NewmanBOOK}.
The values of $w^{\text{part}}_{ji}$ are distributed according to the p.d.f.,
\begin{eqnarray}
&&W^{\text{part}}_q(w^{\text{part}}_{ji})=\int\delta\left(w^{\text{part}}_{ji}-\left(\mu+w^{\text{mf}}_j+\sum_{k=1}^{q-1}w^{\text{eff}}_{jk}\right)\right)\nonumber\\&&\times \left[W^{\text{mf}}_{q}(w^{\text{mf}}_j)\text{d}w^{\text{mf}}_j\right]\prod_{k=1}^{q-1}\left[W^{\text{eff}}(w_{jk}^{\text{eff}})\text{d}w_{jk}^{\text{eff}}\right]~,\label{eq:PartEffectiveFieldDistribution}
\end{eqnarray}
where the sum and product over $k=1,\ldots,q-1$ 
accounts for all neighbours $k$ of cell $j$ except cell $i$.
Eqs.~\eqref{eq:RecursiveEquation} and~\eqref{eq:PartEffectiveFieldDistribution} are coupled and can be solved self-consistently.

The distribution of $\rho_i$ for a given cell $i$ can be written in terms of the distributions of $w_{jk}^{\text{eff}}$ and of $w^{\text{mf}}_j$ using Eqs.~\eqref{eq:FreeEnergyOneSite}, \eqref{eq:HamiltonianOneSite} and~\eqref{eq:Rho},
\begin{eqnarray}
&&R(\rho_i)=\sum_{q}\int\frac{N_q}{N}\nonumber\\&&\times\delta\left\{\rho_i-\left[1+\exp\left(-\beta w^{\text{mf}}_i-\beta\sum_{j=1}^qw^{\text{eff}}_{ij}-\beta\mu\right)\right]^{-1}\right\}\nonumber\\
&&\times\left[W^{\text{mf}}_q(w^{\text{mf}}_i)\text{d}w^{\text{mf}}_i\right]\prod_{j=1}^q\left[W^{\text{eff}}(w_{ij}^{\text{eff}})\text{d}w_{ij}^{\text{eff}}\right]~,\label{eq:rhoDistribution}
\end{eqnarray}
where the sum $j=1,\ldots,q$ is over all neighbours of cell $i$.
The mean of the free energy per cell can be calculated similarly using Eqs.~\eqref{eq:FreeEnergyOneSite}-\eqref{eq:FreeEnergy},
\begin{widetext}
\begin{eqnarray}
{\overline F}/N &=&\sum_{q=1}^{q_{\text{max}}}\frac{N_q}{N}(q-1)\int F^+\left(-\beta w^{\text{mf}}_i-\beta\mu-\beta\sum_{j=1}^qw^{\text{eff}}_{ij}\right)\left[W^{\text{mf}}_q(w^{\text{mf}}_i)\text{d}w^{\text{mf}}_i\right]\prod_{j=1}^q\left[W^{\text{eff}}(w^{\text{eff}}_{ij})\text{d}w^{\text{eff}}_{ij}\right]\nonumber\\
&-&\frac{\overline q}{2}\sum_{q=1}^{q_{\text{max}}}\sum_{q^\prime=1}^{q_{\text{max}}}{\tilde p}_{q}{\tilde p}_{q^\prime}
\int \left[\beta w^{\text{part}}_{ij}+\beta w^{\text{part}}_{ji}+F^+(F^+(\beta w^{\text{part}}_{ij})+F^+(\beta w^{\text{part}}_{ji})+F^-(\beta w^{\text{ff}}_{ij}))+F^-(\beta w^{\text{ff}}_{ij})\right]\nonumber\\&&\times \left[W^{\text{ff}}(w^{\text{ff}}_{ij})\text{d}w^{\text{ff}}_{ij}\right]\left[W^{\text{part}}_q(w^{\text{part}}_{ij})\text{d}w^{\text{part}}_{ij}\right]\left[W^{\text{part}}_{q^\prime}(w^{\text{part}}_{ji})\text{d}w^{\text{part}}_{ji}\right]~.\label{eq:MeanFreeEnergy}
\end{eqnarray}
\end{widetext}
where ${\overline q}=\sum_{q=1}^{q_{\text{max}}}qN_q/N$ is the mean coordination number.
As required, Eqs.~\eqref{eq:rhoDistribution} and~\eqref{eq:MeanFreeEnergy} give the distribution $R(\rho_i)$ and the values of ${\overline F}/N$ in terms of the p.d.f. $W^{\text{eff}}(w_{ij}^{\text{eff}})$, which is defined by Eqs.~\eqref{eq:RecursiveEquation} and~\eqref{eq:PartEffectiveFieldDistribution}.

In order to find the mean free energy per cell, ${\overline F}/N$, and the distribution of mean occupations, $R(\rho_i)$, the distribution, $W^{\text{eff}}(w^{\text{eff}}_{ij})$, of effective fields is determined numerically using Eqs.~\eqref{eq:RecursiveEquation} and~\eqref{eq:PartEffectiveFieldDistribution} in the following way.
First, some initial trial function, $W^{\text{eff}}_0(w^{\text{eff}}_{ij})$, is substituted into Eq.~\eqref{eq:PartEffectiveFieldDistribution} to obtain the distributions, $W^{\text{part}}_q(w^{\text{part}}_{ji})$.
These distributions are then used in Eq.~\eqref{eq:RecursiveEquation} to obtain a new function $W^{\text{eff}}_1(w^{\text{eff}}_{ij})$. 
After this, $W^{\text{eff}}_1(w^{\text{eff}}_{ij})$ is recursively passed to Eqs.~\eqref{eq:RecursiveEquation} and~\eqref{eq:PartEffectiveFieldDistribution} in the same way to obtain a function $W^{\text{eff}}_2(w^{\text{eff}}_{ij})$.
This process is repeated until a function is found which is invariant to the recursive equation within the desired numerical precision.
The final function, $W^{\text{eff}}(w^{\text{eff}}_{ij})$, can then be substituted into Eqs.~\eqref{eq:rhoDistribution} and~\eqref{eq:MeanFreeEnergy} to reveal the required results for ${\overline F}/N$ and $R(\rho_i)$.
The technical details of the numerical solution of Eqs.~\eqref{eq:RecursiveEquation} and~\eqref{eq:PartEffectiveFieldDistribution} are given in App.~\ref{sec:NumericalAppendix}.

In general, there can be one or more stable solutions to the self-consistent Eqs.~\eqref{eq:RecursiveEquation} and~\eqref{eq:PartEffectiveFieldDistribution} which can be found by starting the above iterative procedure with different trial functions $W^{\text{eff}}_0(w^{\text{eff}}_{ij})$.
Each solution has a corresponding mean free energy given by Eq.~\eqref{eq:MeanFreeEnergy}, and the solution with the lowest free energy is assumed to represent the stable state of the system, while the other solutions represent states which are metastable~\cite{LDLandau1980,LandauBinderBOOK}.
It is expected that all of these states can also be reached by dynamical simulations using, e.g. the Metropolis algorithm (see Sec.~\ref{sec:Metropolis}).

The analytical model permits an exact solution only for a particular system topology, i.e.
a random graph with fixed degree distribution and heterogeneous interaction strength.
However, the pore space of real systems may exhibit a different, Euclidean, topology containing many short loops.
Such loops are exponentially rare in a random graph.
A natural question is whether such loops alter the above picture of sorption or not.
Our analysis (see Sec.~\ref{sec:SBA-15}) suggests that the loops in pore space for a structural  model of important porous materials such as SBA-15 may affect the sorption behaviour at low enough temperatures but are insignificant in many situations.

In the next four sections, we first analyse the exact solutions for homogeneous systems (Sec.~\ref{sec:zero-disorder}) and then investigate the effect of disorder in coordination number (Sec.~\ref{sec:CoordDis}), disorder in matrix-fluid and in fluid-fluid interaction strengths (Sec.~\ref{sec:DisorderInteractions}). 
Finally, we analyse the effect of combined disorder of all three types in Sec.~\ref{sec:AllDisorder}.

\section{Results for zero disorder}\label{sec:zero-disorder}

\begin{figure}[ht] 
\includegraphics[width=8.6cm,clip=true]{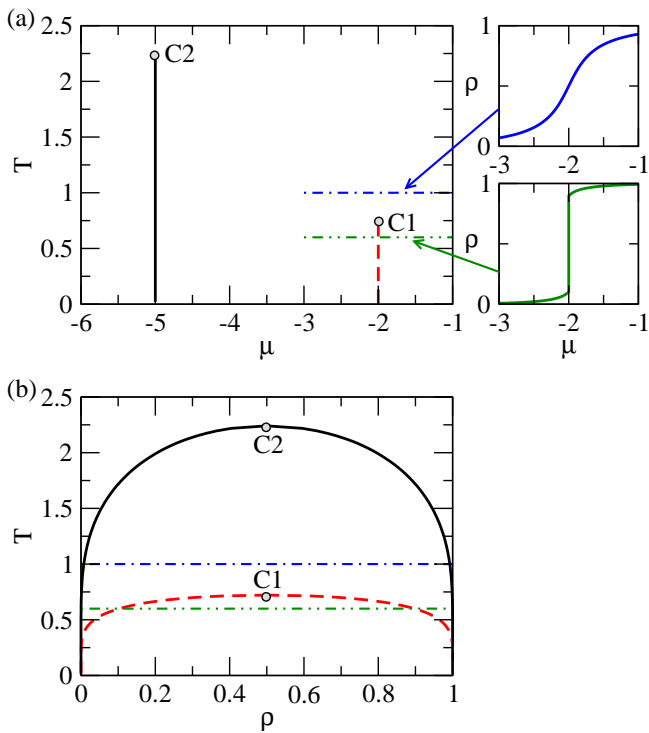}
\caption{ Equilibrium phase diagrams for homogeneous $q$-regular graphs with interaction parameters $w_0^\text{ff}=1$ and $w_0^\text{mf}=0$. (a) $\mu$-$T$ phase diagram for $q$-regular graphs with coordination numbers $q_1=4$ (dashed line) and $q_2=10$ (solid line). 
For both values of $q$, first-order phase transitions occur when crossing the vertical lines below the critical points, C1 for $q_1=4$ at $\mu_1=-2$ and C2 for $q_2=10$ at $\mu_2=-5$. 
The two additional figures in panel (a) show the dependence of the mean density, $\bar{\rho}$, on chemical potential, $\mu$, for a $4$-regular graph above ($T=1.0$, upper figure) and below ($T=0.6$, lower figure) the critical point. 
(b) Liquid-gas coexistence curves (binodals) in the $(\bar{\rho},T)$ plane for fluids in $4$- (dashed line) and $10$-regular graphs (continuous line).
The dot-dashed lines and double dot-dashed lines in both panels correspond to the temperatures for which the density profiles are shown in the right hand figures of panel (a).
\label{fig:Homogeneous}}
\end{figure}

In the absence of any kind of disorder, all the cells have the same coordination number, $q$, and the interaction parameters $w^{\text{mf}}$ and $w^{\text{ff}}$ have $\delta$-functional distributions given by $W^{\text{mf}}_q(w^{\text{mf}})=\delta(w^{\text{mf}}-w^{\text{mf}}_0(q))$ and $W^{\text{ff}}(w^{\text{ff}})=\delta(w^{\text{ff}}-w^{\text{ff}}_0)$, respectively.
In this case, the system can be described by the known equations for a Bethe lattice~\cite{Baxter_Book}, 
so the effective fields (see Eq.~\eqref{eq:EffField}) and partial local fields (see Eq.~\eqref{eq:PartField}) also have $\delta$-functional distributions, $W^{\text{eff}}(w^{\text{eff}}_{ij})=\delta(w^{\text{eff}}_{ij}-w^{\text{eff}}_0)$ and $W^{\text{part}}(w^{\text{part}}_{ij})=\delta(w^{\text{part}}_{ij}-w^{\text{part}}_0)$. 
For concreteness, the energy scale of interactions is set by $w^{\text{ff}}_0=1$.
Under these hypotheses, the problem can be solved exactly~\cite{Baxter_Book} as shown in App.~\ref{sec:ExactSolution}. 
The behaviour is that expected for a classical fluid system~\cite{LDLandau1980}. 
For temperature less than a critical temperature, 
\begin{equation}
T_c=\frac{w^{\text{ff}}_0}{2\ln[q/(q-2)]}~,\label{eq:ExactCritical}
\end{equation}
the system exhibits a first-order phase transition where a high-density phase (liquid) coexists with a low-density phase (gas). 
Fig.~\ref{fig:Homogeneous} shows the phase diagram for two values of the coordination number, $q$. 
As can be seen from the figure, the phase boundary follows a line of constant chemical potential in the space of $T$ and $\mu$, where 
$\mu=-qw^{\text{ff}}/2-w^{\text{mf}}_0$.
This expression for $\mu$ can easily be derived from the mapping of the zero-disorder system to the Ising model. 
As seen from Fig.~\ref{fig:Homogeneous}, the coordination number affects the value of the critical temperature $T_{\text{C}}$ and the chemical potential at the first-order transition (i.e. $T_{\text{C1}}<T_{\text{C2}}$ where $T_{\text{C1}}$ and $T_{\text{C2}}$ are the critical temperatures for $q_1$- and $q_2$-regular graphs, respectively, with $q_2>q_1$) but does not affect the shape of the phase boundary. 
In contrast, we show below that both disorder in coordination number and interaction strengths have a significant impact on the phase diagrams.

\section{Disorder in coordination number}
\label{sec:CoordDis}

In this section, we study the effect of disorder in coordination number in the absence of any other kind of disorder, i.e. the $\delta$-functional distributions of interaction strengths, $W^{\text{mf}}_q(w^{\text{mf}})$ and $W^{\text{ff}}(w^{\text{ff}})$, considered as in Sec.~\ref{sec:zero-disorder}, are used here.
Let us consider a binary system of randomly connected cells with two different coordination numbers, $q=q_1 \ge 3$ and $q=q_2>q_1$, with the number of each equal to, $N_1=fN$ and $N_{2}=(1-f)N$. 
A large difference in coordination numbers $q_1$ and $q_2$ is expected to enhance the effect of disorder in coordination number on sorption.
For this reason, in what follows, we used, for concreteness, $q_1=4$ and $q_2=10$.

\subsection{Sorption and phase diagram for a representative value of $f$}\label{sec:Fixed_f}

\begin{figure} 
\includegraphics[width=8.6cm,clip=true]{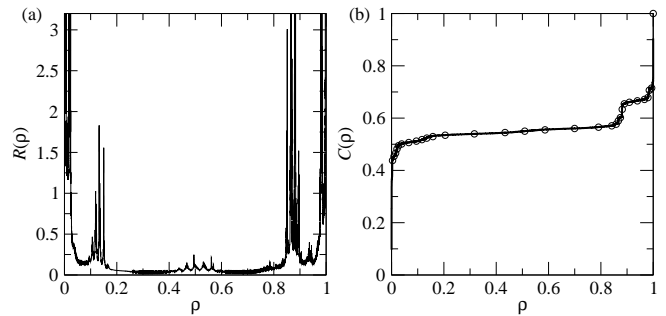}
\caption{Panel (a): Distribution, $R(\rho)$, of cell fluid densities, $\rho$, for binary system with a representative set of parameters, $\mu=-2.8$, $T=0.1$, $w^{\text{ff}}_0=1$ and $w^{\text{mf}}_0=0$, for disorder in coordination number with $f=0.77$. 
The mean value of the density for this distribution is ${\overline \rho}\simeq 0.441$ (see the star in Fig.~\ref{fig:CloseSorptionCurve}(b)).
Panel (b): Cumulative distribution $C(\rho)=\int_{0}^\rho R(\rho)\text{d}\rho$ of the cell fluid densities.
In both panels, the lines are calculated by solving Eqs.~\eqref{eq:RecursiveEquation} and~\eqref{eq:PartEffectiveFieldDistribution} self-consistently and using Eq.~\eqref{eq:rhoDistribution}.
The circles in panel (b) are calculated using Metropolis dynamics simulations described in Sec.~\ref{sec:Metropolis}. The system size is $N=10^6$ and
the mean occupation, $\rho_i$, was measured over a period of $10^5$ Monte-Carlo Steps per Spin (MCSS).\label{fig:DistributionCoordDis}}
\end{figure}

Before analysing the combined effect of $T$ and $f$, it is illustrative to analyse the behaviour of the binary system for a representative choice of parameters $f=0.77$ and $w^{\text{mf}}_0(q_1)=w^{\text{mf}}_0(q_2)=0$.
In order to investigate the phase diagram of this system in the $(\mu,T)$ plane, we explore the sorption curves which give the dependence of the mean value of density, $\overline\rho$, on the chemical potential, $\mu$.
The mean value of density, $\overline\rho$, given by Eq.~\eqref{eq:Mean-rho_def} in terms of the distribution $R(\rho)$, can be calculated using Eqs.~\eqref{eq:rhoDistribution} and~\eqref{eq:Rho}, with the distribution of densities found by solving Eqs.~\eqref{eq:RecursiveEquation} and~\eqref{eq:PartEffectiveFieldDistribution}.
A typical solution for $R(\rho)$ in the case of disorder in coordination number is shown in Fig.~\ref{fig:DistributionCoordDis}.
For the set of parameters used in Fig.~\ref{fig:DistributionCoordDis}, there are two major peaks at $\rho=0$ and $1$, accompanied by many satellite peaks at intermediate densities.
The peaks have a hierarchical structure with several groups associated with different combinations of  
the coordination numbers of neighbouring cells. 
Each group is split into subgroups, corresponding to the coordination numbers of the further neighbours of the cell etc.
For example, the peak at $\rho=0$ corresponds to unoccupied $4$-coordinated cells surrounded by other $4$-coordinated cells.
The structure of $R(\rho)$ appears to be multifractal, similar to that found for the random-field Ising model in one dimension with bimodal disorder in random fields~\cite{Bak1982,Nowotny2001A,Nowotny2001B}.
We have undertaken multifractal analysis using the gliding box method~\cite{PastorSatorras1996}, and have demonstrated that $R(\rho)$ is, in fact, a compact function.
This finding is not surprising as a continuous function $R(\rho)$ has been revealed in all previous analysis on tree-like systems~\cite{BruinsmaPRB1984,NowotnyPRE2002,Sokolovskii2003}.

\begin{figure*} 
\includegraphics[clip=true,width=12.5cm]{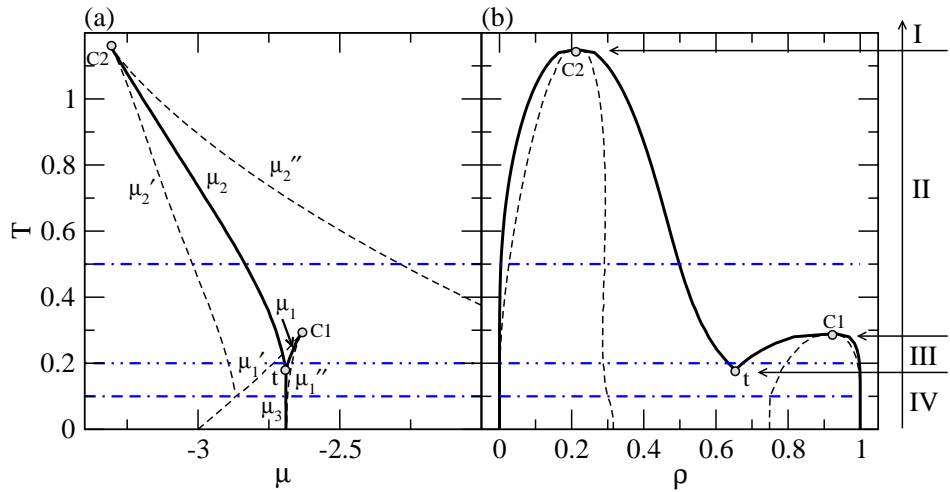}
\caption{Phase diagram in the $(\mu,T)$ plane (a) and binodal in the $({\overline \rho},T)$ plane (b) of a random graph without disorder in interaction strengths and with bimodal disorder in coordination number consisting of a fraction $f=0.77$ of $q_1=4$-coordinated cells and a fraction $1-f$ of $q_2=10$-coordinated cells.
The locations of critical points (C1 and C2) and triple point (t) are marked by circles. 
Solid lines correspond to the coexistence curves of two phases 
where the chemical potential takes values $\mu_1(T)$ between C1 and t, $\mu_2(T)$ between C2 and t, and $\mu_3(T)$ below t. 
The dashed lines correspond to the spinodals, i.e. the boundaries within which metastable states exist, and are marked by $\mu_1^\prime$, $\mu_1^{\prime\prime}$, $\mu_2^\prime$ and $\mu_2^{\prime\prime}$ in panel (a). 
The horizontal dot-dashed, double-dot dashed and double-dash dotted lines correspond to the values of $T$ and $\mu$ for which the sorption curves are shown in Figs.~\ref{fig:RhoMuHighT}(b), \ref{fig:RhoMuHighT}(c) and~\ref{fig:RhoMuHighT}(d), respectively.
The temperature ranges corresponding to the four regimes I, II, III and IV are indicated by the numbers and arrows on the right.
\label{fig:PhaseDiagramf0c77}}
\end{figure*}

\begin{figure} 
\includegraphics[clip=true,width=8.5cm]{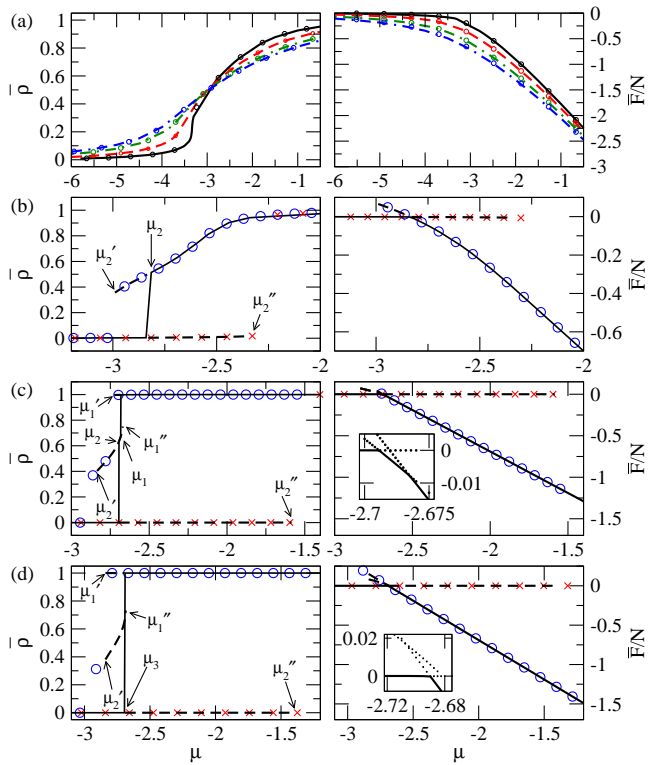}
\caption{Mean fluid density (left panels) and free energy per cell (right panels) as a function of chemical potential for a range of temperatures, $T$. 
The system in all panels has no disorder in interaction strengths and coordination number disorder with $q_1=4$ and $q_2=10$ for $f=0.77$ as in Fig.~\ref{fig:PhaseDiagramf0c77}(a).
Curves represent the substitution of solutions of Eqs.~\eqref{eq:RecursiveEquation} and~\eqref{eq:PartEffectiveFieldDistribution} into Eqs.~\eqref{eq:rhoDistribution} and~\eqref{eq:MeanFreeEnergy}.
Panel (a) shows $\overline\rho$ and ${\overline F}/N$ for several temperatures in regime I (i.e. for $T>T_{\text{C}2}$): $T=1.2$ (solid), $1.5$ (dashed), $1.8$ (dot-dashed) and $2$ (dot double-dashed). Panels (b), (c) and (d) show examples of regime II ($T=0.5$), regime III ($T=0.2$) and regime IV ($T=0.1$), respectively (see dot-dashed, double-dot dashed and double-dash dotted lines in Figs.~\ref{fig:PhaseDiagramf0c77}(a) and~\ref{fig:PhaseDiagramf0c77}(b)).
In panels (b)-(d), the solid lines represent stable states while dashed lines represent metastable states.
The insets in the right hand sides of panels (c) and (d) show a magnification of the free energies of the low- intermediate- and high-density states around the point where they are equal with the stable state marked black and the metastable states marked with dotted lines.
Symbols represent the results of Metropolis dynamics simulations (see Sec.~\ref{sec:Metropolis}). 
The rates of change of $\mu$ are ${\dot\mu}=\pm 10^{-3}\text{MCSS}^{-1}$ (panel (a)) and ${\dot\mu}=\pm 2\times 10^{-5}\text{MCSS}^{-1}$ (panels (b), (c) and (c)) with crosses and circles corresponding to positive, $\dot\mu>0$, and negative, $\dot\mu<0$, rates of change, respectively.
Magnifications of the left hand sides of panels (c) and (d) are shown in Fig.~\ref{fig:CloseSorptionCurve}.
\label{fig:RhoMuHighT}}
\end{figure}

The $\mu$-$T$ phase diagram and the binodal graph corresponding to a system with $f=0.77$ are shown in Figs.~\ref{fig:PhaseDiagramf0c77}(a) and~\ref{fig:PhaseDiagramf0c77}(b). 
Comparison of Figs.~\ref{fig:Homogeneous}(a) and~\ref{fig:Homogeneous}(b) with Figs.~\ref{fig:PhaseDiagramf0c77}(a) and~\ref{fig:PhaseDiagramf0c77}(b) reveals significant differences between the phase diagram and binodals for a homogeneous $q$-regular graph and that for a system with binary disorder in coordination number. 
In particular, the system with binary disorder can exhibit two critical points (C1 and C2) and a triple point (t) instead of a single critical point (C1 or C2) observed in the homogeneous system. 
This leads to a 'Y'-shaped phase diagram (Fig.~\ref{fig:PhaseDiagramf0c77}(a)) and a two-peaked binodal (Fig.~\ref{fig:PhaseDiagramf0c77}(b)) indicative of sorption curves with two steps between three phases.
The binodal features a kink at $T_{\text{t}}$, characteristic of a triple point where three phases can coexist in equilibrium \cite{LDLandau1980}. 
For the choice $f=0.77$, the sorption isotherms display qualitatively different behaviour in four temperature regimes: regime I ($T>T_{\text{C}2}$), regime II ($T_{\text{C}1}<T<T_{\text{C}2}$), regime III ($T_{\text{t}}<T<T_{\text{C}1}$) and regime IV ($T<T_\text{t}$). 
Examples of sorption isotherms for each of these regimes are shown by different panels in Fig.~\ref{fig:RhoMuHighT}.

\begin{figure} 
\includegraphics[width=8cm,clip=true]{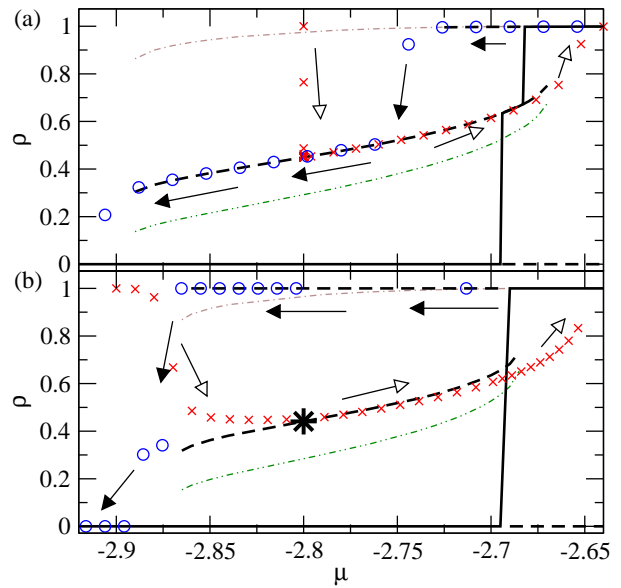}
\caption{Detail of sorption curves of $\overline\rho$ {\it vs} $\mu$ for the same system as in Fig.~\ref{fig:RhoMuHighT} and $T=0.2$ (panel (a)) and $T=0.1$ (panel (b)). 
Solid and dashed lines represent stable and metastable states as in Figs.~\ref{fig:RhoMuHighT}(c) and~\ref{fig:RhoMuHighT}(d). 
The values of $\overline\rho$ for the intermediate-density states averaged across only $q_1$- and $q_2$-coordinated cells are shown by double-dot dashed and dot-dashed lines, respectively.
Symbols represent Metropolis dynamics simulations for rates of change of $\mu$ given by $\dot\mu=2\times 10^{-5}$ (panel (a)) and $\dot\mu=10^{-6}$ (panel (b)) (same symbol styles as in Fig.~\ref{fig:RhoMuHighT}). 
Black arrows indicate the variation of $\mu$ in numerical simulations of desorption from the stable state with $\rho \sim 1$ (density shown by circles). 
White arrows show the variation of the system from a metastable state with $\rho \sim 1$ (density shown by crosses).
The star refers to the mean density of the p.d.f. presented in Fig.~\ref{fig:DistributionCoordDis}(a).
\label{fig:CloseSorptionCurve}}
\end{figure}

In regime I, the equilibrium value of $\overline\rho$ varies smoothly with $\mu$, not showing any features (see solid and dashed lines at progressively reduced temperatures in Fig.~\ref{fig:RhoMuHighT}(a)).
In regime II, the value of ${\overline\rho}(\mu)$ (see solid line in Fig.~\ref{fig:RhoMuHighT}(b)) exhibits a first-order phase transition at some value of chemical potential, $\mu=\mu_2$, i.e. there is a vertical jump in the sorption curve at $\mu_2$. 
As $\mu$ passes through $\mu_2$, most of the cells with a higher coordination number suddenly change state, i.e. in case of adsorption (desorption), most of the $q_2$-coordinated cells become occupied (empty).
For temperatures close to $T\alt T_{\text{C}2}$, the size of the jump approaches zero continuously, i.e. transition between regimes I and II is continuous.
In Fig.~\ref{fig:RhoMuHighT}(b), the solid and dashed curves refer to solutions of Eqs.~\eqref{eq:RecursiveEquation} and~\eqref{eq:PartEffectiveFieldDistribution}, corresponding to stable and metastable states of the system, respectively.
For values of $\mu$ in the range $\mu^\prime_2<\mu<\mu^{\prime\prime}_2$ (the region between the dashed spinodal lines in Fig.~\ref{fig:PhaseDiagramf0c77}(a)), there are two solutions, with mean densities $\rho_1(\mu)$ and $\rho_2(\mu)>\rho_1(\mu)$, corresponding to two possible phases of the system, of low and intermediate density. 
In the first low-density phase, most cells are unoccupied. 
In the second intermediate-density phase, the $q_2$-coordinated cells are mostly occupied while the $q_1$-coordinated cells are mostly unoccupied. 
The chemical potential at the first-order phase transition in regime II takes the value $\mu_2(T)$ which can be found from the condition, $F_1(\mu_{2})=F_2(\mu_{2})$, for coexistence of the intermediate- and low-density phases. 
The free energies $F_1(\mu)$ and $F_2(\mu)$ can be calculated using Eq.~\eqref{eq:MeanFreeEnergy} and are presented in the right panel of Fig.~\ref{fig:RhoMuHighT}(b).

In regime III, there are two phase boundaries (see Fig.~\ref{fig:PhaseDiagramf0c77}(a)) and a third phase can exist in addition to the low- and intermediate-density phases observed in regime II. 
In the new phase, most cells in the system are occupied and therefore the density, $\rho_3(\mu)$, is higher than in the other two phases (see upper branch in Fig.~\ref{fig:RhoMuHighT}(c)). 
For values of the chemical potential, $\mu$, falling simultaneously between the spinodal lines $\mu^\prime_1(T)<\mu<\mu^{\prime\prime}_1(T)$ and between the spinodal lines $\mu^\prime_2(T)<\mu<\mu^{\prime\prime}_2(T)$ (see dashed spinodal lines in Fig.~\ref{fig:PhaseDiagramf0c77}(a)) all three phases can be observed.
For such $\mu$, one of the phases is the stable state of the system, while the other two phases are metastable states.
The intermediate- and high-density phases can coexist at $\mu=\mu_1$ when their free energies are the same,
i.e. $F_3(\mu_1)=F_2(\mu_1)$ (see Fig.~\ref{fig:RhoMuHighT}(c) right panel).
As a consequence, two jumps can appear in the sorption curve (see solid curve in Fig.~\ref{fig:RhoMuHighT}(c) and the magnification of the double step in Fig.~\ref{fig:CloseSorptionCurve}(a)),
corresponding to the coexistence either of the low- and intermediate-density phases (at $\mu_2$) or of the intermediate- and high-density phases (at $\mu_1$).
At each jump, sorption occurs in cells of a particular coordination number, with the jump at lower $\mu=\mu_2$ corresponding to sorption in the $q_2$-coordinated cells 
and at $\mu=\mu_1>\mu_{2}$ corresponding to sorption in the $q_1$-coordinated cells. 

The intermediate density phase in the two step regime exists for $\mu_2^\prime<\mu<\mu_1^{\prime\prime}$.
However, it is stable only in a relatively narrow interval, $\mu_2<\mu<\mu_1$ (see Fig.~\ref{fig:CloseSorptionCurve}(a)), and metastable outside this range (see dashed line at intermediate values of $\rho$ in Fig.~\ref{fig:CloseSorptionCurve}(a)).
The nature of the intermediate density phase can be clarified by calculating the relative densities of $q_1$ (see dashed curve in Figs.~\ref{fig:CloseSorptionCurve}(a) and (b)) and $q_2$ (see dot-dashed curve in Figs.~\ref{fig:CloseSorptionCurve}(a) and (b)) coordinated cells.
It follows from Fig.~\ref{fig:CloseSorptionCurve}(a) that in the intermediate state, practically all of the $q_2$-coordinated cells are occupied, while a much smaller fraction of the $q_1$-coordinated cells are occupied, given that the mean density is in a range $0.4\alt {\overline\rho}\alt 0.8$.

As the temperature approaches the triple point from above $T\agt T_{\text{t}}$, the values of $\mu_1$ and $\mu_2$ corresponding to the two jumps in the sorption curve become closer together, i.e. $\mu_2\simeq \mu_1$, and they merge with each other at $T=T_{\text{t}}$.
The triple point thus corresponds to the coexistence of all three phases when they all have the same free energy, $F_1(\mu_{\text{t}})=F_2(\mu_{\text{t}})=F_3(\mu_{\text{t}})$.

Regime IV corresponds to $T<T_\text{t}$ where there is a single jump in which cells of both coordination numbers become occupied simultaneously.
This jump occurs at $\mu=\mu_{3}$ (see Fig.~\ref{fig:RhoMuHighT}(d)) and corresponds to coexistence of the low- and high-density phases, with $F_1(\mu_{\text{t}})=F_3(\mu_{\text{t}})$.
At such temperatures, the intermediate-density phase is not observed in equilibrium since either $F_2(\mu)>F_1(\mu)$ or $F_2(\mu)>F_3(\mu)$ for all values of $\mu$ (see Fig.~\ref{fig:CloseSorptionCurve}(b) and the inset of Fig.~\ref{fig:RhoMuHighT}(d)).
However, the second phase does exist as a metastable state (see the middle dashed curve in Fig.~\ref{fig:RhoMuHighT}(d) and its magnification in Fig.~\ref{fig:CloseSorptionCurve}(b)) of the system and can be observed in a non-equilibrium situation (see Sec.~\ref{sec:Metropolis} for more detail).

The metastable states within each of the four temperature regimes can also be analysed in the $\overline\rho$-$T$ plane (see Fig..~\ref{fig:PhaseDiagramf0c77}(b)).
Indeed, the two peaked dashed lines in Fig.~\ref{fig:PhaseDiagramf0c77}(b) represent the spinodals corresponding to the first order phase transitions between low- and intermediate-density phases (left peak) and between the intermediate- and high-density phases (right peak).
Any points below the binodal but not below either of the spinodals represents a metastable state, while points within the two spinodals represent states which are unstable to spinodal decomposition.

\subsection{Dependence on $f$}\label{sec:Dependence_f}

\begin{figure*}[ht] 
\includegraphics[width=17cm,clip=true]{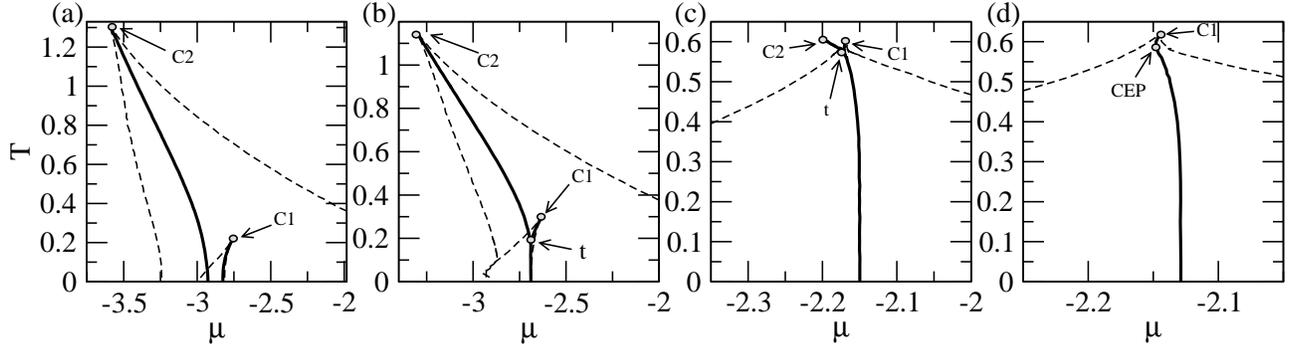}
\caption{Phase diagrams for a system consisting of a mixture of $q_1=4$ and $q_2=10$ coordinated nodes and constant interaction strengths.
The fraction of $q_1$-coordinated nodes is (a) $f=0.7$ (b) $f=0.77$ (c) $f=0.95$ and (d) $f=0.957$.
Locations of lines of first-order phase transitions (solid lines) and boundaries of the regions where metastable states exist (dashed lines) are shown in the space of chemical potential and temperature.
The locations of critical points (C1 and C2), critical end points (CEP) and triple points (t) are marked with circles for the values of $f$ where they exist.
The critical end point is unique and is found at $T\simeq 0.586$, $\mu\simeq -2.1481$ and $f\simeq 0.957$.
The first-order transition lines are found by using Eqs.~\eqref{eq:RecursiveEquation}-\eqref{eq:MeanFreeEnergy}.
\label{fig:PhaseCoordDis}}
\end{figure*}

The $\mu$-$T$ phase diagram is highly influenced by the relative fraction of cells with different coordination number.
Fig.~\ref{fig:Homogeneous}(a) together with the sequence of Figs.~\ref{fig:PhaseCoordDis}(a)-(d) show the evolution of the phase diagram as the fraction $f$ of low-coordinated ($q_1$) cells is increased from $0$ to $1$.
As shown below, there are three characteristic values, $f_1$, $f_{\text{t}}$ and $f_2$, at which the phase diagram experiences qualitative changes.

\begin{figure}[ht] 
\includegraphics[width=5cm,clip=true]{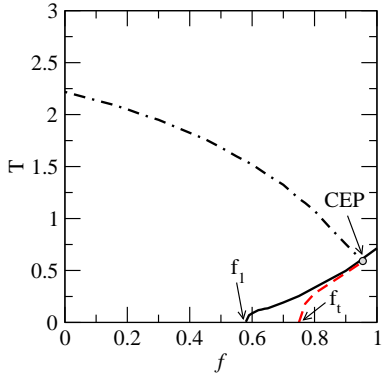}
\caption{Values of tricritical temperature $T_{\text{t}}$ (dashed line) and critical temperatures $T_{\text{C}1}$ (solid line) and $T_{\text{C}2}$ (dot-dashed line) as a function of fraction $f$ of $q_1=4$-coordinated nodes.
The value of $f_2$ corresponds to the location of the critical end point (marked CEP). \label{fig:CritTempVary_CoNumDis_q4-10}}
\end{figure}

When $f=0$, the system is a $q_2$-regular graph and the phase diagram shown in Fig.~\ref{fig:Homogeneous}(a) (continuous line) is recovered. 
Similarly, the phase diagram for a pure $q_1$-regular graph is obtained for the extreme case with $f=1$ (see dashed line in Fig.~\ref{fig:Homogeneous}(a)).
In general, the value of $\mu_2(T,f)$ depends on both $T$ and $f$ unless $f=0$ or $f=1$ in which case it becomes independent of temperature (see Fig.~\ref{fig:Homogeneous}).
The value of $T_{\text{C}2}$ decreases with growing $f$ (see dot-dashed line in Fig.~\ref{fig:CritTempVary_CoNumDis_q4-10}). 
When $f<f_1$, the phase transition at $\mu_2(T,f)$ is dominated by the transformation of cells with high coordination, $q_2$. 
Accordingly, sorption isotherms can either be continuous if $T>T_{\text{C}2}$ (regime I) or exhibit a single discontinuity if $T<T_{\text{C}2}$ (regime II).
When $f>f_1$, a second phase boundary, describing a first-order transition for $q_1$-coordinated cells, can be found at $\mu=\mu_1(T,f)$ and temperatures $T<T_{\text{C}1}(f)$ (see Fig.~\ref{fig:PhaseCoordDis}(a)). 
The critical point, $T_{\text{C}1}(f)$, is located at zero temperature for $f=f_1$, i.e. $T_{\text{C}1}(f_1)=0$, and it increases with increasing $f$ (see solid line in Fig.~\ref{fig:CritTempVary_CoNumDis_q4-10}). 

As the fraction $f$ increases further in the region $f_1<f<f_{\text{t}}$, the separation between the two phase boundaries $\mu_1(T,f)$ and $\mu_2(T,f)$ at zero temperature reduces until they merge when $f=f_{\text{t}}$, and for $f_{\text{t}}<f<f_2$ there is a triple point, marked by $t$ in Figs.~\ref{fig:PhaseCoordDis}(b) and~\ref{fig:PhaseCoordDis}(c). 
The temperature, $T_{\text{t}}$, at which the triple point occurs increases with increasing values of $f$ (see dashed line in Fig.~\ref{fig:CritTempVary_CoNumDis_q4-10}). 
Sorption isotherms of systems with $f$ in the interval $f_t<f<f_2$ have a rich behaviour that can be in any of the four regimes described in detail in Sec.~\ref{sec:Fixed_f}.
For large enough values of $f$, the value of $T_{\text{C}2}$ becomes smaller than $T_{\text{C}1}$ and there is a new regime, $T_{\text{C}2}<T<T_{\text{C}1}$, in which only the first-order phase transition in the $q_1$-coordinated nodes occurs (see Fig.~\ref{fig:PhaseCoordDis}(c)).
In this regime of concentrations, $f_{\text{t}}<f<f_2$, and at zero temperature, it has been found numerically that in the ground state all cells are in the same state as each other (either occupied or unoccupied corresponding to the low- and high-density phases mentioned in the discussion of regime IV illustrated in Fig.~\ref{fig:RhoMuHighT}(c)).
The value of $\mu_3(T=0,f)$ at the phase boundary between the two ground states can then be calculated 
by equating to each other the values of the internal energies, $U(\{\tau_i=1\})$ and $U(\{\tau_i=0\})$ of the fully occupied and fully unoccupied phases, respectively, given by,
\begin{eqnarray}
U(\{\tau_i=1\})&=&\sum_{q=1}^{q_{\text{max}}}N_q\left(\frac{qw^{\text{ff}}_0}{2}+\mu\right)~\text{and}\nonumber\\
U(\{\tau_i=0\})&=&0~.
\end{eqnarray}
This gives the value of the critical chemical potential at zero-temperature as,
\begin{equation}
\mu_3(T=0,f)=\frac{{\overline q}w^{\text{ff}}_0}{2}~.\label{eq:ZeroTempTransition}
\end{equation}
In other regimes for $f$, Eq.~\eqref{eq:ZeroTempTransition} does not necessarily hold, because 
the system in the ground state can contain both occupied and unoccupied cells,
and thus the estimate for $\mu_3(T=0,f)$ becomes harder to find~\cite{BruinsmaPRB1984,Middleton2001,Liu2007}.

When the fraction $f=f_2\simeq 0.957$, the critical point C2 merges with the triple point, t, and a critical end point (marked by CEP in Fig.~\ref{fig:PhaseCoordDis}(e) and Fig.~\ref{fig:CritTempVary_CoNumDis_q4-10}) is observed.
Notice that the critical temperature $T_{\text{C}1}$ is slightly higher than the values of $T_{\text{C}2}$ and $T_{\text{t}}$ at the critical end point (the solid line passes above the merge point of the dashed and dot-dashed lines in Fig.~\ref{fig:CritTempVary_CoNumDis_q4-10} and the point marked C1 on Fig.~\ref{fig:PhaseCoordDis}(d) is above the point marked CEP).
For $f>f_2$, there is only a single first-order phase transition, with a phase boundary $\mu_1(f,T)$ and a critical point $T=T_{\text{C}1}$.

\subsection{Role of matrix-fluid interaction}\label{sec:matrix-fluid}

Above, we analysed the case when $w_0^{\text{mf}}(q_1)=w_0^{\text{mf}}(q_2)=0$ and the value of $w^{\text{ff}}$ is constant.
However, in a real system, the matrix-fluid interaction strength can depend on coordination number.
The simplest form of such a dependence could be represented by the following linear relation, $w^{\text{mf}}_0(q)=(q_{\text{max}}-q)w^{\text{mf}}_1+{\tilde w}^{\text{mf}}_0$, where $w^{\text{mf}}_1$ and ${\tilde w}^{\text{mf}}_0$ do not depend on $q$ and $q_{\text{max}}$ is the largest coordination number present in the lattice.
The linear term in this equation could represent the increased interaction between fluid in cells of low coordination number and the matrix.
Indeed, the coordination number in our model describes the number of fluid neighbours of a cell.
The surface area of a cell can be split into fluid-fluid contact area and matrix-fluid contact area.
Assuming that the surface areas of all cells are equal, the greater the fluid-fluid contact area (i.e. the coordination number) the smaller the matrix-fluid contact area, i.e. the strength of matrix-fluid interaction.

In the analysis above, $w^{\text{mf}}_1={\tilde w}^{\text{mf}}_0=0$. 
The picture does not change qualitatively when ${\tilde w}^{\text{mf}}_0\neq 0$ and $w^{\text{mf}}_1=0$. Indeed, the value of $w_0^{\text{mf}}(q)$ is independent of $q$ and the phase diagram in Fig.~\ref{fig:PhaseCoordDis} only exhibits a parallel shift along the $\mu$ axis. 
In contrast, if $w_0^{\text{mf}}(q_1)\neq w_0^{\text{mf}}(q_2)$, $w^{\text{mf}}_1\neq 0$, a different form of phase diagram can be observed.
For example, if $w_0^{\text{mf}}(q_1)+q_1w^{\text{ff}}_0/2= w_0^{\text{mf}}(q_2)+q_2w^{\text{ff}}_0/2=-\mu_0$, the Hamiltonian is invariant under the transformation $\tau_i\to 1-\tau_i$, $\mu-\mu_0\to -(\mu-\mu_0)$ and thus shows the same symmetry as the Ising model.
This is because for that choice of parameters, the Hamiltonian can be written as an Ising Hamiltonian, 
${\cal H}=-(w^{\text{ff}}/4)\sum_{\langle ij\rangle}s_is_j-(\mu-\mu_0)\sum_{i}s_i/2+\text{const}$, where $s_i=2\tau_i-1$.
The phase diagram for this case is known and the critical temperature can be calculated exactly~\cite{Barrat2008,Dorogovtsev2002,Dorogovtsev2008,DorogovtsevREVIEW,CohenBOOK}.
Another choice for the matrix-fluid interaction strength is 
$w_0^{\text{mf}}(q_1)+q_1w^{\text{ff}}= w_0^{\text{mf}}(q_2)+q_2w^{\text{ff}}=-2\mu_0$.
The phase diagrams for this case is the same as those shown in Fig.~\ref{fig:PhaseCoordDis}, reflected about the line $\mu=\mu_0$.
\begin{figure} 
\includegraphics[width=8.6cm,clip=true]{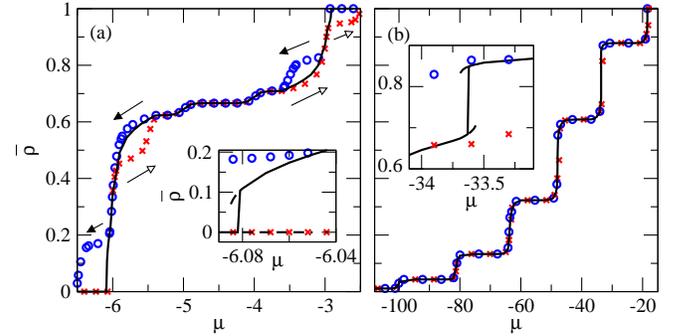}
\caption{Sorption curves for systems with different interaction strengths for cells of different coordination number. 
Panel (a): $q_1=3$, $q_2=4$, $f=0.666$, $w^{\text{mf}}_1=5$ ${\tilde w}^{\text{mf}}_0=0$ and $T=0.05$. 
The inset magnifies the region of the low $\mu$ first order phase transition at $\mu\simeq -6.08$.
Panel (b): Binomial distribution of coordination numbers with $q_{\text{max}}=20$, $p=0.9$, $w^{\text{mf}}_1=20$, ${\tilde w}^{\text{mf}}_0=0$ and $T=0.05$. 
The inset shows the detail of the region where one of the first order phase transitions occurs at $\mu\simeq -33.6$.
The crosses (adsorption) and circles (desorption) refer to the results of Metropolis dynamics with $\dot\mu=10^{-5}\text{MCSS}^{-1}$ (panel (a)) and $\dot\mu=10^{-4}\text{MCSS}^{-1}$ (panel (b)). \label{fig:VariableMF}}
\end{figure}

In the general case with non-zero values of the parameters $w^{\text{mf}}_1$ and ${\tilde w}^{\text{mf}}_0$, the picture can be very rich.
Consider, for example a binary system consisting of a fraction $f=0.666$ of cells with coordination number $q_1=3$ and the rest with coordination number $q_2=4$. 
For interaction parameters $w^{\text{mf}}_1=5$ and ${\tilde w}^{\text{mf}}_0=0$ and temperature $T=0.05$, the system exhibits two phase transitions located at $\mu\simeq -6.08$ and $\mu\simeq -2.91$, where vertical jumps in $\overline\rho$ occur (see Fig.~\ref{fig:VariableMF}(a) with a magnification of the lower jump in the inset).
These jumps correspond to a macroscopic change in the fraction of occupied $q_2$- and $q_1$-coordinated cells, respectively.
The jumps are accompanied by the presence of metastable states, shown for one of the jumps by the dashed lines in the inset in Fig.~\ref{fig:VariableMF}(a).

The use of the linear form for $w^{\text{mf}}_0(q)$ described above and of a binomial distribution of coordination numbers $N_q=N{\binom{q_{\text{max}}}{q}}p^q(1-p)^{q_{\text{max}}-q}$ produces a system which is equivalent to that studied in Ref.~\cite{Sokolovskii2003}.
If the values of the parameters are chosen to be $w^{\text{mf}}_1=20$, $q_{\text{max}}=20$ and $p=0.9$ (in the notation of Ref.~\cite{Sokolovskii2003} this corresponds to $I=1$, $K=20$ and $p_1=0.9$)
then the system exhibits two first-order phase transitions in a certain range of temperatures (see solid line in Fig.~\ref{fig:VariableMF}(b)).
These transitions occur at $\mu\simeq -33.6$ and $\mu\simeq -18.42$ and correspond to the two highest steps in the ``staircase'' in Fig.~\ref{fig:VariableMF}(b).
A magnification of the second highest step is shown in the inset of Fig.~\ref{fig:VariableMF}(b).
Similarly to the case shown in Fig.~\ref{fig:VariableMF}(a), there are metastable states near the phase transition.
Each step in Fig.~\ref{fig:VariableMF}(b) corresponds to sorption in cells of a given coordination number.
The phase transition is observed only when the cells of that coordination number percolate and when the system has low enough disorder, i.e. at high values of $\mu$, the only unoccupied cells are those of coordination number $q=20$, meaning that the system is effectively less disordered than at lower values of $\mu$.
In other words, possible phase transitions for low $q\le 18$-coordinated cells are destroyed by disorder.

\section{Disorder in interaction strengths}\label{sec:DisorderInteractions}

\begin{figure} 
\includegraphics[width=7cm,clip=true]{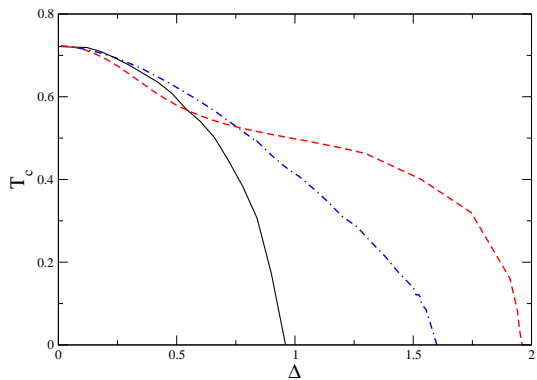}
\caption{Critical temperatures as a function of degree of disorder for a $4$-regular graph with normal (solid curve) and exponential (dot-dashed) disorder in $w^{\text{mf}}$ (i.e. $\Delta^{\text{mf}}=\Delta$, $\Delta^{\text{ff}}=0$) and cut-normal distribution in $w^{\text{ff}}$ ($\Delta^{\text{ff}}=\Delta$, $\Delta^{\text{mf}}=0$; dashed curve).\label{fig:CriticalTemp_Disorder}}
\end{figure}

In this section, we discuss the effect of disorder in the strengths for both matrix-fluid and fluid-fluid interactions.
The matrix surrounding a given cell is inherently heterogeneous. 
This heterogeneity can be of a chemical nature, due to roughness or due to the shape of the pore, e.g. its diameter.
The matrix-fluid interaction strength depends on this heterogeneity, and can result in two types of disorder,
symmetric disorder, which can be described by a normal distribution with p.d.f., $W^{\text{mf}}(w^{\text{mf}})=N(w^{\text{mf}}_0,\Delta^{\text{mf}})$, and asymmetric disorder, represented below by an exponential distribution, $W^{\text{mf}}(w^{\text{mf}})=\exp(-(w^{\text{mf}}-w^{\text{mf}}_0)/\Delta^{\text{mf}}-1)$ for $w^{\text{mf}}>w^{\text{mf}}_0-\Delta^{\text{mf}}$ and $W^{\text{mf}}(w^{\text{mf}})=0$, otherwise.
Symmetric disorder originates from small scale disorder due to chemical heterogeneity and roughness.
Asymmetry in the p.d.f. might occur when there is disorder in the shape of the pores~\cite{Handford2013b}.
An expected symmetric peak-shaped distribution of the fluid-fluid interaction is approximately modelled below by a modified normal distribution which is cut so that $w^{\text{ff}}$ cannot be negative, i.e. $W^{\text{ff}}(w^{\text{ff}})\propto \exp\left(-\left(w^{\text{ff}}-w^{\text{ff}}_0\right)^2/2\left({\Delta^{\text{mf}}}\right)^2\right)$ for $w^{\text{ff}}>0$ and $W^{\text{ff}}(w^{\text{ff}})=0$ when $w^{\text{ff}}\le 0$.

First, we analyse the effect of the disorder on the critical behaviour of the system.
Introducing disorder reduces the critical temperatures, $T_{\text{C}}(\Delta^{\text{mf}})$ and $T_{\text{C}}(\Delta^{\text{ff}})$, of the second-order phase transition gradually from its value, $T_{\text{C}}(\Delta=0)$, found at zero disorder.
As disorder increases, the critical temperatures go to zero at different critical values of the disorder, $\Delta_{\text{C}}^{\text{mf}}$ and $\Delta_{\text{C}}^{\text{ff}}$, in matrix-fluid and fluid-fluid interaction strengths, respectively (see Fig.~\ref{fig:CriticalTemp_Disorder}).
This behaviour of the critical temperature is similar to that found for the random field Ising model with a normal distribution of random fields~\cite{Cardy1996}.
This is expected because the lattice gas model with disorder in matrix-fluid and fluid-fluid interactions can be mapped to the random-field Ising model and to the Ising model with correlated disorder in fields and bonds, respectively~\cite{Kierlik1998}.

\begin{figure} 
\includegraphics[width=8.6cm,clip=true]{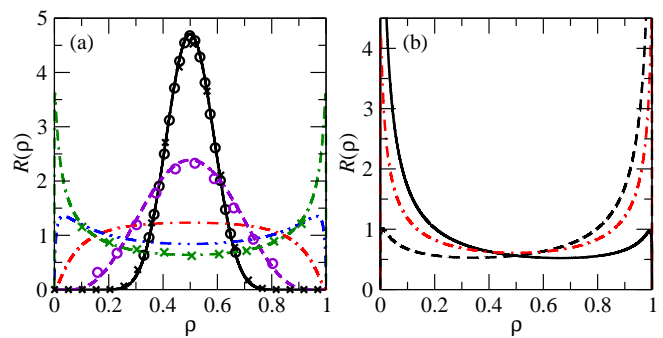}
\caption{Distributions of the mean occupancies, $\rho_i$, of the cells in a $q=4$-regular graph with $\mu=-2$, constant $w^{\text{ff}}$ for all cells and normal disorder in $w^{\text{mf}}$.
Panel (a): Temperature, $T=1.0$ and width of disorder $\Delta^{\text{mf}}=0.3$ (solid curve), $0.6$ (dashed), $1.2$ (dot-dashed), $1.8$ (double-dot dashed) and $2.4$ (dot double-dashed).
The circles represent a normal distribution of width given by Eq.~\eqref{eq:RhoDistributionApproxWidth} and the crosses were found by Metropolis dynamics simulations in which the values of $\rho_i$ for each of $N=10^6$ cells were averaged over a duration of $10^3$~MCSS.
Panel (b): $T=0.4$ with $\Delta^{\text{mf}}=0.75$ (solid and dashed curves representing low- and high-density states, respectively) and $0.8$ (dot-dashed curve). \label{fig:DistributionsInterDis}}
\end{figure}

A prominent characteristic of sorption 
in a disordered system is that the mean fluid densities $\rho_i$ take different values depending on the neighbourhood of cell $i$, and are distributed with the p.d.f. $R(\rho_i)$.
Let us consider the case of normally distributed disorder in $w^{\text{mf}}$, when the temperature is greater than the critical temperature, $T>T_{\text{C}}(\Delta=0)$, for the zero-disorder case. 
For such temperatures, at any value of $\Delta^{\text{mf}}$ the sorption curve for the mean density, $\overline \rho$, is a continuous and monotonically growing function of $\mu$, i.e. it does not exhibit any discontinuities characteristic of phase transitions.
However, the distribution of $\rho_i$ exhibits an interesting evolution with degree of disorder.
When the degree of disorder is small, the distribution $R(\rho_i)$ is approximately normal (see App.~\ref{sec:ExactSolution}), with the mean $\overline \rho$ corresponding approximately to the mean for zero disorder (see solid and dashed lines in Fig.~\ref{fig:DistributionsInterDis}(a)).
When $\Delta^{\text{mf}}$ takes intermediate values, the distribution is approximately uniform (double-dot dashed line).
For larger disorder, the distribution becomes bimodal-like, with two peaks at $\rho_i\simeq 0$ and $\rho_i\simeq 1$ (dot double-dashed line in Fig.~\ref{fig:DistributionsInterDis}(a)).
In this disorder-controlled regime~\cite{Handford2013b}, 
most cells in the system have become either occupied or unoccupied and do fluctuate from these states.
For very large disorder and $\mu=-qw^{\text{ff}}/2$ (corresponding to ${\overline\rho}=1/2$), fluid in approximately half of the cells has a large repulsive interaction with its surroundings and thus the cells are unoccupied, while fluid in the rest of the cells has an strong attractive interaction with the matrix, and these cells are occupied.
This disorder controlled regime is also observed (not shown) for exponential disorder in matrix-fluid interaction and disorder in fluid-fluid interactions. 

We have also analysed the distribution $R(\rho_i)$ for the system with normally distributed disorder in matrix-fluid interactions below the critical temperature for zero disorder, $T<T_{\text{C}}(\Delta=0)$.
For such values of disorder, at a particular value of chemical potential $\mu$, both low- and high-density stable phases can coexist.
The distribution of $R(\rho_i)$ in both of these two phases can feature
one prominent peak and one smaller secondary peak near $\rho\simeq 0$ and $1$ (see solid and dashed lines in Fig.~\ref{fig:DistributionsInterDis}(b)).
As $\Delta^{\text{mf}}$ approaches the critical value from below, each of these curves transforms so that the peaks are more equal in size,
and exactly at the value of disorder, $\Delta^{\text{mf}}=\Delta^{\text{mf}}_{\text{C}}(T)$, at which $T_{\text{C}}(\Delta^{\text{mf}})=T$, they merge continuously into the same curve (shown by dot-dashed line in Fig.~\ref{fig:DistributionsInterDis}(b)).
For $\Delta^{\text{mf}}>\Delta^{\text{mf}}_{\text{C}}(T)$, there is a single stable phase characterised by a single distribution of $R(\rho_i)$ with two peaks (see dashed line in Fig.~\ref{fig:DistributionsInterDis}(b)), which depends on $\Delta^{\text{mf}}$ in a similar way to the curves presented in Fig.~\ref{fig:DistributionsInterDis}(a) for higher temperature.

Such an evolution of the distribution of densities for $T<T_c(\Delta=0)$ can be understood in a qualitative way as follows. 
Similarly to the high temperature case, there are some cells in the system where fluid has a large attractive or repulsive interaction with its surrounding matrix.
The state of these cells is determined by the quenched disorder and is not influenced by thermal fluctuations.
In addition, some cells have an intermediate interaction with the matrix, meaning that their states are set by the states of the surrounding cells. 
If the concentration of these cells is great enough they can form a percolating cluster and the system can have two global states depending upon whether cells within this cluster are occupied or unoccupied (see also Ref.~\cite{Swift1994}).
With increasing disorder, the relative number of disorder-controlled cells increases, meaning that the intermediate cells cease to be able to percolate at $\Delta^{\text{mf}}_{\text{C}}(T)$ and the system moves to the disorder-controlled regime for $\Delta^{\text{mf}}>\Delta^{\text{mf}}_{\text{C}}(T=0)$.

\section{Disorder in coordination number and interaction strengths}\label{sec:AllDisorder}

\begin{figure} 
\includegraphics[width=6cm,clip=true]{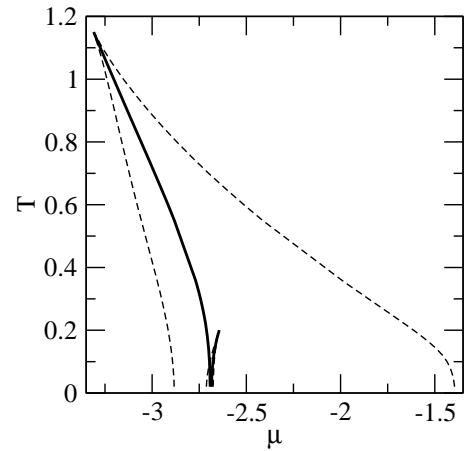}
\caption{Phase diagram for binary system of $q_1=4$ and $q_2=10$-coordinated cells with a fraction $f=0.77$ of $q_1$-coordinated cells.
The matrix-fluid interactions are normally distributed with coordination number dependent standard deviations, $\Delta^{\text{mf}}(q=4)=0.24$ and $\Delta^{\text{mf}}(q=10)=0$, while disorder in fluid-fluid interactions is a cut-normal distribution with $\Delta^{\text{ff}}=0.04$. \label{fig:AllDisPhaseDiagram}}
\end{figure}

The solution for the mean density and free energy presented in Sec.~\ref{sec:Solution} can be applied to a system with disorder in coordination number, $q$, and both interaction strengths $w^{\text{mf}}$ and $w^{\text{ff}}$.
In the case when disorder in interaction strength is not significant, i.e. $\Delta^{\text{mf}}\ll {\overline {w^{\text{ff}}}}$ and $\Delta^{\text{ff}}\ll {\overline {w^{\text{ff}}}}$, the dependence of the locations of the phase boundaries on the fraction, $f$, of $q_1$-coordinated cells is qualitatively the same as that shown in Fig.~\ref{fig:PhaseDiagramf0c77}(a). 
An example of the phase diagram in this regime is shown in Fig.~\ref{fig:AllDisPhaseDiagram}.
It is found that the critical temperatures $T_{\text{C}1}$, $T_{\text{C}2}$ and the temperature of the triple point $T_{\text{t}}$ are all shifted to lower values relative to those for zero disorder in interaction strengths (c.f. Fig.~\ref{fig:AllDisPhaseDiagram} and Fig.~\ref{fig:PhaseDiagramf0c77}(a)).
In the opposite case of large disorder in either $w^{\text{mf}}$ or $w^{\text{ff}}$, no phase transitions are observed at any temperature, i.e. all long range order is lost and the picture is identical to that discussed in Sec.~\ref{sec:DisorderInteractions}.

\section{Metropolis Dynamics Simulations}\label{sec:Metropolis}

In order to test the results of the analytical model, we
have used Metropolis dynamics simulations~\cite{Metropolis1953,LandauBinderBOOK} for the evolution of the system at a given temperature and both fixed and variable values of the chemical potential.
Such simulations involve changing the states of a set of $N$ cells according to the following rules.
Consider the system of cells in a certain state $\{\tau_i\}=\{\tau^\prime_i\}$ with energy ${\cal H}(\{\tau^\prime_i\})$, where $\cal H$ is given by Eq.~\eqref{eq:Hamiltonian}.
In a Monte Carlo step of Metropolis dynamics simulation, a cell is selected at random and its state is changed (either occupied to unoccupied or vice-versa), so that the new state of the system is $\{\tau_i\}=\{\tau^{\prime\prime}_i\}$, and has energy ${\cal H}(\{\tau^{\prime\prime}_i\})$. 
If the energy change, $\Delta{\cal H}={\cal H}(\{\tau^{\prime\prime}_i\})-{\cal H}(\{\tau^\prime_i\})$, is negative, the cell is left in its new state, otherwise it is changed back to its original state with probability $p=1-\exp(-\beta \Delta{\cal H})$.
Time scales within the Metropolis dynamics are measured in terms of Monte-Carlo steps per spin (MCSS), with spin being equivalent to cell.
The efficiency of the standard Metropolis scheme is limited at low temperatures since most transition proposals are rejected (i.e. are null) and nothing changes in the system for long periods of time.
To improve the computational performance of our simulations, we use a level 1 Monte Carlo Adsorbing Markov Chain (MCAMC) method~\cite{LandauBinderBOOK,Novotny1995a,Novotny1995b}, which effectively skips the null transitions (see details in App.~\ref{sec:MetropolisAppendix}).

An important step in the above procedure is the calculation of the internal energy of the system.
The value of ${\cal H}$ depends on the connectivity between pores and between the pores and the matrix.
The strength of interactions between fluid in the cells and the matrix is a site characteristic, and has a random value $w^{\text{mf}}_i$ chosen from the distribution $W_q(w^{\text{mf}}_i)$ which depends on the coordination number.
The strength of interactions between fluid in neighbouring occupied cells is a bond characteristic, and has a distribution $W^{\text{ff}}(w^{\text{ff}})$.
The connections between the cells are defined by the topology of the network.
In our analysis, the topology of a random graph with fixed degree distribution was used~\cite{NewmanBOOK}.
Such a random graph can be generated by creating $N_q$ nodes, each with $q$ ``stub'' bonds, for every value of $q$ in the range $1\le q\le q_{\text{max}}$.
Two stub bonds are then randomly selected and connected together to form a link. 
The process is then repeated until all stubs are connected.

Following the rules of Metropolis dynamics, we can obtain estimates of the distribution, $R(\rho_i)$, of fluid densities at particular values of the chemical potential and temperature (see crosses in Fig.~\ref{fig:DistributionsInterDis}(a) and circles in Fig.~\ref{fig:DistributionCoordDis}(b)) and the mean density $\overline \rho$ of the fluid per cell as a function of changing chemical potential at fixed temperature (see symbols in Figs.~\ref{fig:RhoMuHighT}, \ref{fig:CloseSorptionCurve}, and~\ref{fig:VariableMF}).
For disorder in interaction strengths away from the critical point, the agreement between the distribution $R(\rho_i)$ found by numerical simulation and by analytical calculation is almost perfect.
Near the critical point, the occupied cell-occupied cell correlation length is too high to get good agreement for a finite sized system.
For disorder in coordination number, the Metropolis simulations cannot reveal the details of all of the peaks seen in Fig.~\ref{fig:DistributionCoordDis}(a). 
It can, however, reproduce the cumulative distribution well (c.f. the line and circles in Fig.~\ref{fig:DistributionCoordDis}(b)).

Similarly, the free energy can be calculated using Metropolis dynamics and compared with analytical results.
In order to do this, the free energy, $F_{\pm\infty}$, is found for $\mu=\pm\mu_\infty$ where $\mu_\infty\gg w^{\text{ff}}$ meaning that $\langle\tau_i\rangle\simeq 0$ or $1$ for all $N$ cells.
The free energy is then given by $F=\langle{\cal H}\rangle-TS\simeq {\cal H}(\{\tau_i\})$ with $S$ being the entropy which is equal to zero for a fully occupied or fully empty system.
The value of $F$ at other values of $\mu$ can then be found by monitoring $\rho$ as $\mu$ is slowly increased or reduced and using the formula $F=-\int_{\pm\mu_\infty}^\mu \rho\text{d}\mu+F_{\pm\infty}$~\cite{LandauBinderBOOK}, valid in the range of $\mu$ where $\rho(\mu)$ is a continuous function.
In the right panels of Fig.~\ref{fig:RhoMuHighT}, the free energy calculated in this way is compared with the analytical calculation given by Eq.~\eqref{eq:FreeEnergy}, giving excellent agreement.
It should be noted that, at the location of the spinodals, i.e. at the value of $\mu=\mu^\prime_2$ or $\mu^{\prime\prime}_2$ where the density changes in a step-wise manner, the above formula for the free energy is not applicable. 

Another interesting point to comment on is about how to access metastable states using Metropolis dynamics.
Indeed, our analytical approach presented in Sec.~\ref{sec:Solution} provides a description of both the stable equilibrium state and several metastable states which can be destroyed only by nucleation due to thermal fluctuations.  
Metropolis dynamics simulations allows all of these states to be attained.
In Fig.~\ref{fig:RhoMuHighT}(c) the stable states are represented by the solid line, while the metastable states are shown by dashed lines.
There are three types of metastable states, corresponding to three possible phases.
Simulations in which $\mu$ is varied slowly enough will always follow the equilibrium state for any given value of $\mu$.
However, in practise this regime is not achievable for a random-graph (a mean-field like system) due to the prohibitively large number of Monte-Carlo steps required for nucleation to occur in certain regions.
Instead, 
it is expected that the relaxation time from the metastable to the stable state diverges exponentially with system size~\cite{Paul1989}, so that 
for increasing or decreasing $\mu$ the simulated system follows first the equilibrium state (solid line) until the first-order transition, and then explores the metastable states (dashed curve).
As the value of $\mu$ approaches $\mu^\prime_{1,2}$ or $\mu^{\prime\prime}_{1,2}$ the energy barrier preventing the system from leaving the metastable state reduces in size and the system is eventually able to move to another phase.
A similar behaviour can be seen for all other temperatures shown in Fig.~\ref{fig:RhoMuHighT} and for both systems shown in Fig.~\ref{fig:VariableMF}.

Four examples of the exploration of the metastable states for the intermediate- and high-density phases at two different temperatures (two for each temperature) are shown by the black and white arrows in Figs.~\ref{fig:CloseSorptionCurve}(a) and~\ref{fig:CloseSorptionCurve}(b). 
In the first two examples (circles and black arrows in both panels), all cells are initially occupied at a high value of $\mu$. 
The value of $\mu$ is gradually reduced at a rate ${\dot\mu}=\text{d}\mu/\text{d}t=-2\times 10^{-5}\text{MCSS}^{-1}$ (panel (a)) and $-2\times 10^{-5}\text{MCSS}^{-1}$ (panel (b)), and the system follows first the stable state (solid curve) and then metastable state (dashed curve) with highest density.
In fact, even after the high-density analytical solution ceases to exist $\mu=\mu^{\prime}_1$ (see the finite range of the dashed line at high density in both panels), the Metropolis dynamics simulation at the rate used does not immediately step to the intermediate-density state.
For lower values of ${\dot\mu}$, the system makes a quicker transition to the intermediate-density state.
As $\mu$ continues to decrease, the system follows the intermediate-density metastable state (dashed line at intermediate values of $\rho$), until that solution also ceases to exist at $\mu=\mu^{\prime}_2$, after which a transition occurs to the low-density state.
In the final two examples of a Metropolis dynamics simulation (crosses and white arrows in both panels), all cells are initially occupied at $\mu=-2.8$ (Fig.~\ref{fig:CloseSorptionCurve}(a)) and $\mu=-2.9$ (Fig.\ref{fig:CloseSorptionCurve}(b)). 
At $T=0.2$ (panel (a)), the system was first allowed to relax for a time $10^3\,$MCSS at a constant value of $\mu$ (see vertical line at $\mu=-2.8$) so that it enters the intermediate-density metastable state.
Next, the value of $\mu$ was gradually increased at a rate ${\dot\mu}=2\times 10^{-5}\,\text{MCSS}^{-1}$.
At $T=0.1$ (panel (b)), the system was prepared into the intermediate-density metastable state by increasing $\mu$ from $-2.9$ to $-2.8$ in a period of $1000\,\text{MCSS}$ (see downward pointing white arrow).
The value of $\mu$ was then slowly increased at a rate ${\dot\mu}=10^{-6}\,\text{MCSS}^{-1}$.
The state of the system achieved in Metropolis dynamics simulations follows the intermediate-density metastable state (dashed line), passing through the range of $\mu$ corresponding to the stable state (solid line) and remaining in the intermediate-density phase until it no longer exists (see right-hand limit of the dashed line at intermediate values of $\rho$ at $\mu=\mu^{\prime\prime}_1$).
After $\mu$ passes $\mu^{\prime\prime}_1$, the density continuously increases until the system is in the high-density stable state.
A more rapid upward jump in $\rho$ can be seen by reducing $\dot\mu$. 

\section{Random-graph model for SBA-15}
\label{sec:SBA-15} 

The analytical model developed in the previous sections is valid for a random graph topology with local tree-like structure (i.e. with an exponentially small number of short loops). 
Real porous media such as mesoporous silicas, aerogels, Vycor glass, and others exhibit pore network space which can be strongly interconnected and short loops might be present in these materials~\cite{Horikawa2011,Coasne_13:review}. 
In order to investigate the role of the pore network topology and, in particular, the role of loops on the shape of sorption isotherms, we use a minimal structural model for a typical mesoporous material, such as mesoporous silica SBA-15~\cite{Zhao1998}, and compare the sorption in such a model embedded in Euclidean space with that in the corresponding random-graph model.

\begin{figure}[h] 
\includegraphics[width=10cm,clip=true]{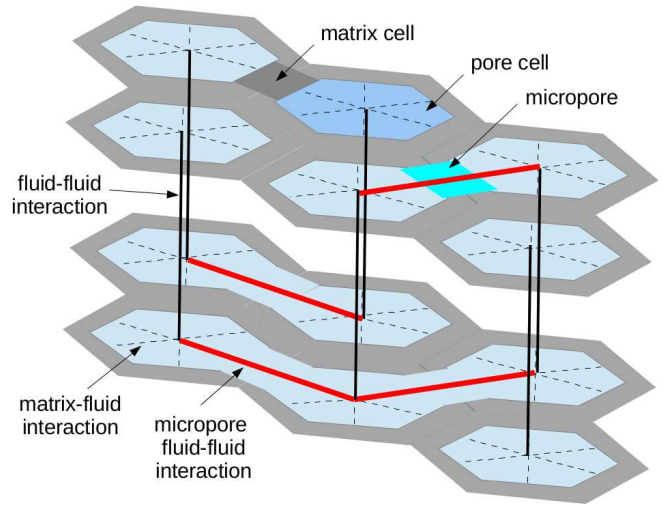}
\caption{ The pore cell network for the Euclidean model of SBA-15. 
The light grey (blue) hexagons represent the pore cells vertically connected to each other by thin solid lines (vertical fluid-fluid interactions, $w^{\text{ff}}_{\text{vert}}$). 
The thick horizontal lines account for horizontal fluid-fluid interactions, $w^{\text{ff}}_{\text{horiz}}$, between pore cells in different vertical mesopores through the micropores. 
The cylindrical walls of the mesopores are split into matrix cells shown by 
dark grey hexagon sides.  
The dashed horizontal lines represent matrix-fluid interactions, $w^{\text{mf}}$. 
 \label{fig:SBA_structure}}
\end{figure}

An ordered mesoporous silica SBA-15 consists of hexagonally arranged cylindrical pores. 
The rough cylinder walls contain micropores which interconnect neighbouring cylindrical mesopores~\cite{Coasne_13:review}. 
We model the structure of SBA-15 by a 3-dimensional lattice of pore cells (see Fig.~\ref{fig:SBA_structure}). 
The lattice consists of horizontal layers of pore cells (2 layer fragments are shown in Fig.~\ref{fig:SBA_structure}). 
The pore cells are vertically linked to each other thus forming cylindrical mesopores. 
The axes of the cylinders form a triangular lattice while the matrix boundaries (represented by dark grey area between the columns) exhibit hexagonal structure. 
Each pore cell in the interior region (not on the surface) of the column  is surrounded by one pore cell from above and one from below. 
Horizontally, each pore cell in the column is surrounded by matrix cells (the sides of hexagons shown in dark grey). 
Due to the presence of micropores a pore cell, $i$, can have  a random  number of neighbouring matrix (silica) cells, $n_{\text{matr}}(i)$, varying from six (no micropores) to zero when the pore cell does not have any matrix neighbours. 
If a matrix cell is absent in the neighbourhood of the pore cell, this means that there is a micropore at this place and pore cell is linked to the pore cell in the neighbouring column. 
Therefore, the coordination number, $q_i$, of the interior pore cell $i$ (number of neighbouring pore cells) is $q_i=8-n_{\text{matr}}(i)$ and thus $q_i=2$ if the pore cell is not communicating to other cylindrical pores through the micropores.   
The location of the micropores on cylindrical walls is assumed to be random and uncorrelated, so that any matrix cell in any vertical cylindrical pore is removed and replaced by a micropore with probability $p_{\text{t}}$  which is a parameter of the model. 
The value of $p_{\text{t}}$ 
can be estimated from the experimentally measured ratio of micropore and mesopore volume found to be of $\alt 10^{-1}$~\cite{Morishige_13} which gives $p_{\text{t}} \alt 10^{-1}$. 
Below, we analyse two representative cases with $p_{\text{t}} = 0.05$ and $p_{\text{t}} = 0.2$. 

The number $n_{\text{matr}}$ is given by a binomial distribution, $\text{Pr}(n_{\text{matr}})=\binom{6}{n_{\text{matr}}}p_{\text{t}}^{6-n_{\text{matr}}}(1-p_{\text{t}})^{n_{\text{matr}}}$, parameterised by $p_{\text{t}}$. 
The probability distribution of the pore cell degree can also be expressed in terms of $p_{\text{t}}$ as $\text{Pr}(q)=\binom{6}{q-2}p_{\text{t}}^{q-2}(1-p_{\text{t}})^{8-q}$ so that the mean cell degree is $\overline{q}=2+6p_{\text{t}}$.
The pore space of the structural model embedded in Euclidean space (Euclidean model, see Fig.~\ref{fig:SBA_structure}) contains loops due to the random transverse micropore links connecting vertical columns. 
The counterpart random-graph model is built from a set of randomly connected nodes with exactly the same realisation of degree distribution as for the  Euclidean model. 
More precisely, 
in a realisation of the Euclidean model all the links are cut resulting in a  set of nodes with $q$ stubs. 
The vertical and horizontal stubs corresponding to vertical and horizontal links in the Euclidean model, respectively, are distinguished in order to account for possible differences between fluid-fluid interactions in the mesopores (vertical) and micropores (horizontal). 
The stubs of the same type are then randomly connected to each other thus making the  random-graph model of the original Euclidean model.

If the pore cells are filled with fluid then they can interact both with the neighbouring matrix and fluid cells. 
The interactions are described by the lattice-gas Hamiltonian given by Eq.~\eqref{eq:Hamiltonian}.  
All the cells in the columns are vertically connected by  links representing fluid-fluid interactions (solid vertical lines in Fig.~\ref{fig:SBA_structure}). 
The horizontal (transverse) dashed lines represent the matrix-fluid interactions between  pore cells  and its neighbouring matrix cells, shown as dark grey sides of the hexagons. 
The transverse  bold solid (red) lines  account for the fluid-fluid interactions between pore cells in different vertical columns occurring through the micropores, i.e. micropore fluid-fluid interactions. 
The strengths of these interactions (parameters of the model) are proportional to the contact surface  area between the cells. 
As known from experiment~\cite{Ryoo_00,Naumov_08,Morishige_13}, the vertical pore diameter is of $\simeq 5$ - $10~$nm, the micropore size  $\alt 3~$nm which is of the same order as the thickness of the pore walls. 
In our model, the horizontal size of the cell coincides with the vertical pore diameter and the vertical size is chosen to be of the same order as the micropore size. 
In this case, the mean strength of the fluid-fluid interactions through the micropores, $w^{\text{ff}}_{\text{horiz}}$, is much smaller than that between the cells in the same cylindrical pore, $w^{\text{ff}}_{\text{vert}}$, i.e. $w^{\text{ff}}_{\text{horiz}}/w^{\text{ff}}_{\text{vert}} \sim 0.1$. 
Both the vertical and horizontal fluid-fluid interactions are assumed to be the same for all the cells, i.e. the variations in contact surface area are ignored for simplicity.  
The matrix-fluid interaction strength per unit area is  of the same order of magnitude as the fluid-fluid interaction per unit area, i.e. $w^{\text{mf}}/w^{\text{ff}}_{\text{horiz}} \agt 1$. 
The inequality sign in the last relation accounts for typically greater attraction between fluid and matrix than between fluid and fluid. 

In order to take into account the roughness of the cylinder walls, the strength of the  matrix-fluid interaction, $w^{\text{mf}}_{ij}$, between the pore cell $i$ and neighbouring matrix cell $j$ is assumed to be an identically, independently and normally distributed random variable with p.d.f., $N(w^{\text{mf}}_0,\Delta^{\text{mf}})$. 
The total strength, $w^{\text{mf}}_{i}=\sum_j^{n_{\text{matr}}(i)}w^{\text{mf}}_{ij}$, of the matrix-fluid  interaction for cell $i$ depends on the number of surrounding matrix cells, $n_{\text{matr}}$, and thus is normally distributed according to 
$N(n_{\text{matr}}(i)w^{\text{mf}}_0,n^{1/2}_{\text{matr}}(i)\Delta^{\text{mf}})$.

\begin{figure}[h] 
\includegraphics[width=8.6cm,clip=true]{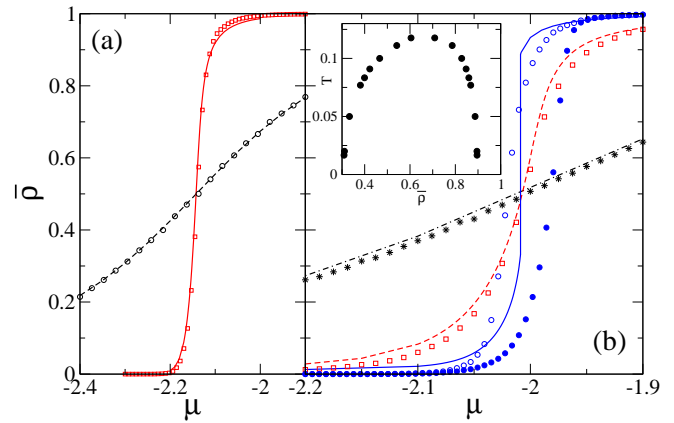}
\caption{Sorption isotherms for structural models of SBA-15 with 
$w^{\text{ff}}_{\text{horiz}}/w^{\text{ff}}_{\text{vert}}=0.1$,  $w^{\text{mf}}_0=0.2$ and $\Delta^{\text{mf}}=0.04$.   
The lines correspond to the results for the random-graph  model. 
The symbols represent the results of Metropolis dynamics simulations for $50\times 50 \times 50$ pore  cells arranged on a lattice (see Fig.~\ref{fig:SBA_structure}) with periodic horizontal and open vertical boundary conditions.
(a)  $p_{\text{t}}=0.05$, $T^{-1}=2$ (dashed line and squares), $T^{-1}=15$ (solid line and circles); 
(b)  $p_{\text{t}}=0.2$, $T^{-1}=2$ (double dot-dashed line and stars),  
$T^{-1}=5$ (dashed line and squares), 
$T^{-1}=20$ (solid line, solid and open circles for adsorption and desorption, respectively, with $\dot{\mu}= 10^{-8}~$MCSS$^{-1}$).  
The inset shows the binodal (phase coexistence curve) for the random-graph model with critical temperature  $T_{\text{c}}\simeq 0.118 \pm 0.002$. 
 \label{fig:SBA15vh_sorption}}
\end{figure}

\begin{figure}[h] 
\includegraphics[width=8.6cm,clip=true]{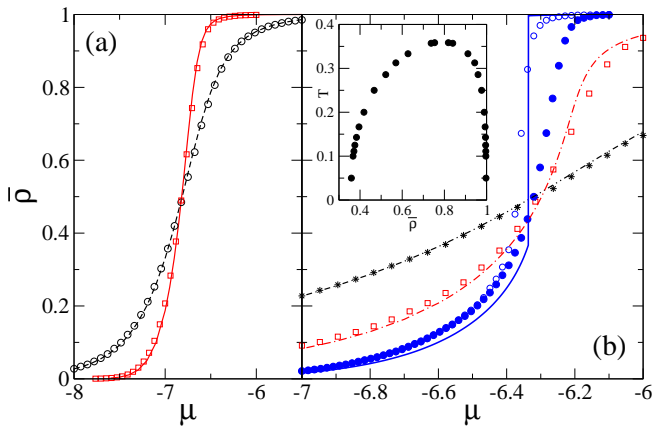}
\caption{Sorption isotherms for structural models of SBA-15 with 
$w^{\text{ff}}_{\text{horiz}}=w^{\text{ff}}_{\text{vert}}= 1$, $w^{\text{mf}}_0=1$ and $\Delta^{\text{mf}}=0.2$. 
  The lines and symbols are for the random-graph and Euclidean models, respectively. 
(a)  $p_{\text{t}}=0.05$, $T^{-1}=2$ (dashed line and circles),  $T^{-1}=10$ (solid line and squares); 
(b)  $p_{\text{t}}=0.2$, $T^{-1}=1$ (double dot-dashed line and stars),  
$T^{-1}=2$ (dot-dashed line and squares), 
$T^{-1}=10$ (solid line, solid and open circles for adsorption and desorption, respectively, with $\dot{\mu}= 10^{-7}~$MCSS$^{-1}$).  
The inset shows the binodal (phase coexistence curve) for the random-graph model with critical temperature $T_{\text{c}}\simeq 0.358\pm 0.001$. 
 \label{fig:SBA15_sorption}}
\end{figure}

The equilibrium sorption isotherms in  Euclidean network model of SBA-15 can be studied using Metropolis dynamics simulations 
(as described in Sec.~\ref{sec:Metropolis}) while for the random-graph model 
 the analytical approach (see Sec.~\ref{sec:Solution}) can be employed. 
The comparison of sorption isotherms for Euclidean and random-graph models for a typical set of parameters and for a particular realisation of node degrees taken from $\text{Pr}(q)$ (configuration averaging does not change the picture discussed below) is shown in Fig.~\ref{fig:SBA15vh_sorption}(a) and (b) for  two values of 
$p_{\text{t}} = 0.05$ ($\overline{q}=2.3$) and $p_{\text{t}} = 0.2$ ($\overline{q}=3.2$), respectively. 
As mentioned in Sec.~\ref{sec:zero-disorder}, the energy scale is set up by fixing $w^{\text{ff}}_{\text{vert}}=1$ and the distributions of the strength of matrix-fluid interactions are chosen to be the same in both models.
It follows from Fig.~\ref{fig:SBA15vh_sorption}  that there is a good agreement between the two models for 
all the temperatures for which the equilibrium in Metropolis dynamics simulations has been achieved, i.e. when there is no macroscopic coexistence between liquid and gas (cf. lines and symbols in Fig.~\ref{fig:SBA15vh_sorption}(a) and (b)). 
The agreement between the curves for the two models demonstrates that the loops in the pore space are insignificant for sorption processes provided liquid-gas coexistence is not involved.
For sufficiently high values  of $p_{\text{t}} \ge p^*_{\text{t}} $ (with $p^*_{\text{t}}\simeq 0.155 \pm 0.005$ for the set of parameters used in Fig.~\ref{fig:SBA15vh_sorption}), the random-graph model predicts a  coexistence curve below a critical temperature 
$T_{\text{c}}$ ($T_{\text{c}}\simeq 0.118 \pm 0.002$ as seen from the inset in Fig.~\ref{fig:SBA15vh_sorption}(b)). 
As the temperature approaches $T_{\text{c}}$, the agreement between sorption isotherms obtained in the random-graph and Euclidean models becomes worse (cf. the dashed line and squares  in Fig.~\ref{fig:SBA15vh_sorption}(b)). 
This can be ascribed to (i) the lack of equilibration of numerical simulations and (ii) the effect of pore space loops.  
Indeed, below the critical temperature, the random-graph model exhibits a  first-order phase transition  (see the vertical jump for the solid curve in   Fig.~\ref{fig:SBA15vh_sorption}(b)). 
In the same temperature range, the Metropolis dynamics used for the Euclidean model shows significant slowing down and the system ceases to equilibrate within the available simulation time. 
This results in different adsorption (solid circles) and desorption (open circles) isotherms which can be only tentatively compared with the equilibrium 
random-graph isotherm. 

The contribution of the pore space loops to the discrepancy between sorption isotherms for random-graph and Euclidean models found near and below the critical temperature can be understood in a qualitative way. 
The density of fluid in porous media varies  with $\mu$ in a series of  abrupt changes which represent avalanches in absorbed 
fluid~\cite{DetcheverryPRE2005}. 
Above the critical temperature, most of these avalanches are  small  in comparison with the typical size of pore space loops. 
The loops are therefore not important for sorption processes. 
In contrast, around and below $T_{\text{c}}$, avalanches of all sizes including a  macroscopic avalanche responsible for the vertical step in density occur in the system (see the solid line in Figs.~\ref{fig:SBA15vh_sorption}(b) and~\ref{fig:SBA15_sorption}(b)). 
Therefore, the loops in the pore space can influence the dynamics of condensation and affect the shape of the sorption curves.   

In order to explore the effect of the parameters of the model, we have undertaken a similar analysis for  different sets of parameters. 
In particular, Fig.~\ref{fig:SBA15_sorption} shows the results for systems with the same interaction strength for vertical and horizontal fluid-fluid interactions (i.e. $w^{\text{ff}}_{\text{horiz}}=w^{\text{ff}}_{\text{vert}}$).
All the main qualitative features are the same as in Fig.~\ref{fig:SBA15vh_sorption}(a)-(b). As expected, the quantitative characteristics are different, e.g. $T_{\text{c}}\simeq 0.358\pm 0.001$ and $p_{\text{t}}^*\simeq 0.138 \pm 0.005$. 

Overall, we found that the random-graph model serves as a good representation of the Euclidean network model for pore space in mesoporous silica SBA-15 and mimic well the equilibrium sorption isotherms for 
temperatures above the critical point appearing at sufficiently high concentration of micropores. 
For temperatures below the critical point, the random-graph model becomes less reliable. 
However, it predicts a discontinuity of the mean density in the same range of chemical potential where the hysteresis loop exists for the non-equilibrium Metropolis dynamics simulations in Euclidean network. 
It should be noted that the minimal structural model of SBA-15 used above does not take into account the gradual layer adsorption on internal surface of the cylindrical pores which is responsible for the initial increasing part of the isotherms found experimentally. 
However, the random-graph model reveals the significance of the network features of the pore space in SBA-15 and predicts the existence of a critical concentration of  micropores above which the coexistence of two phases is possible in SBA-15.

\section{Conclusions}\label{sec:Conclusions}

To conclude, we have presented an exact solution for the equilibrium distribution of densities and mean free energy of a heterogeneous lattice gas condensation model on a random graph of fixed degree distribution.
This solution is given in the form of two coupled integral equations which can be solved numerically using discrete Fourier transforms.
The solution is based on and follows from known solutions obtained by recursive techniques for the random field Ising model on a Bethe lattice~\cite{BruinsmaPRB1984}.
Our approach allows disorder in both coordination number and interaction strengths to be considered.
The solution of these integral equations for all types of disorder has been supported by independent numerical simulations of sorption using Metropolis dynamics.
An advantage of the analytical solution is that we are able to identify and analyse both stable and metastable states, even in regions which are challenging to access by numerical approaches.

Disorder in coordination number can lead to interesting physical phenomena.
Namely, in a random graph with a mixture of cells of two greatly different coordination numbers, multiple phase transitions exist.
At a given temperature and an increasing chemical potential, first the subsystem of cells with high coordination number becomes occupied due to the large number of interacting neighbours at each such cell, while the rest of the system remains unoccupied.
A further increase in chemical potential leads to a second phase transition in which adsorption occurs in the rest of the cells.
Such a multiple phase transition might be related to wetting and filling transitions seen in numerical simulations~\cite{Page1996,Alvarez1999,Sarkisov2000,Brennan2002}.
Disorder in interaction strengths is found to reduce the critical temperature of the phase transition, with the critical temperature reaching zero at a certain value of disorder which corresponds to a zero-temperature critical disorder.
For large enough disorder in interaction strengths, the cells with repulsive (attractive) interactions with matrix are unoccupied (occupied), thus producing a distribution of densities with two peaks.


We have also demonstrated that the random-graph model mimics the sorption well in realistic Euclidean  models of porous media, such as SBA-15, at temperatures above the critical one. 
This gives an opportunity to use exactly solvable  random-graph models to  study quantitatively capillary condensation at high enough temperature (or in porous media with few topological loops) and provide qualitative insight at low temperatures (or media with relatively high densities of loops).

\section{Acknowledgement}  

T.P.H. acknowledges the financial support of UK EPSRC. 


\appendix

\section{Numerical Solution of Self-Consistent Equations}\label{sec:NumericalAppendix}

The solution, $W^{\text{eff}}(w^{\text{eff}}_{ij})$, of the coupled integral Eqs.~\eqref{eq:RecursiveEquation} and~\eqref{eq:PartEffectiveFieldDistribution} has previously been found both by discretizing this function~\cite{Sokolovskii2003} and by means of an approach using population dynamics~\cite{Swift1994,Mezard2001}. 
We use the former method, which can be performed in the following way.

In order to discretize the functions $W^{\text{eff}}_s(w^{\text{eff}}_{ij})$, we note that they can be non-zero only in a certain region $0\le w^{\text{eff}}<w^{\text{eff}}_{\text{max}}=w^{\text{ff}}_{\text{max}}$, where $w^{\text{ff}}_{\text{max}}$ is the largest possible strength of the fluid-fluid interaction.
Therefore, they can be represented numerically by 
storing their values at a series of $M=2^{17}$ points, $W_{s,m}=W^{\text{eff}}_s(m\Delta)$ ($0\le m<M$) where $\Delta=x_{\text{max}}/M$ with some value $x_{\text{max}}>w^{\text{eff}}_{\text{max}}$.
In order to interpolate the value of this function between stored values, we calculate the first $M/2$ components of its Fourier series expansion, with the highest frequency $(2\Delta)^{-1}$, so that
the numerical representation of $W^{\text{eff}}_s(w^{\text{eff}}_{ij})$ is given by,
\begin{equation}
W^{\text{eff}}_{s,\text{num}}(w^{\text{eff}})=\sum_{n=-M/2+1}^{M/2}{\tilde W}_{s,n}\exp\left(2\pi in\frac{w^{\text{eff}}}{x_{\text{max}}}\right)~,\label{eq:WeffFourier}
\end{equation}
where ${\tilde W}_{s,n}$ are defined by a discrete Fourier transform (DFT),
\begin{equation}
{\tilde W}_{s,n}=M^{-1}\sum_{m=0}^{M-1}{W}_{s,m}\exp\left(-\frac{2\pi inm}{M}\right)~.
\end{equation}

In order to iteratively solve Eqs.~\eqref{eq:RecursiveEquation} and~\eqref{eq:PartEffectiveFieldDistribution}, we find a series of successively improved approximations $W^{\text{eff}}_s(w^{\text{eff}}_{ij})$ to the function $W^{\text{eff}}(w^{\text{eff}})$.
We begin the iterative method by substituting the distribution $W^{\text{eff}}_{s,\text{num}}(w^{\text{eff}})$ into Eq.~\eqref{eq:PartEffectiveFieldDistribution}.
This gives the function $W^{\text{part}}_q(w^{\text{part}}_{ji})$, which is the convolution of $W^{\text{eff}}_{s,\text{num}}(w^{\text{eff}})$ with itself $q-1$ times and with $W^{\text{mf}}_q(w^{\text{mf}}_i)$ once. 
Such a convolution can be performed in Fourier space, giving the first $M/2$ terms in the Fourier series of $W^{\text{part}}_q(w^{\text{part}}_{ji})$ as,
\begin{equation}
{\tilde W}^{\text{part}}_{q,n}=x_{\text{max}}^{q-1}{\tilde W}_{s,n}^{q-1}{\tilde W}^{\text{mf}}_n~,\label{eq:WpartWeff}
\end{equation}
where ${\tilde W}^{\text{mf}}_n=x^{-1}_{\text{max}}\int_{w^{\text{mf}}_{\text{min}}}^{w^{\text{mf}}_{\text{max}}}W^{\text{mf}}_q(w^{\text{mf}}_{j})\exp(-2\pi in(w^{\text{mf}}_{j}-w^{\text{mf}}_{\text{min}})/x_{\text{max}})\text{d}w^{\text{mf}}_{j}$ is the Fourier series of $W^{\text{mf}}_q(w^{\text{mf}}_i)$, with $w^{\text{mf}}_{\text{max}}$ and $w^{\text{mf}}_{\text{max}}$ being the maximum and minimum non-zero values of the matrix-fluid interaction, respectively. 
The numerical representation of the function $W^{\text{part}}_q(w^{\text{part}}_{ji})$ is then given by,
\begin{eqnarray}
&&W^{\text{part}}_q(w^{\text{part}}_{ji})=\nonumber\\
&&\sum_{n=-M/2+1}^{M/2}{\tilde W}^{\text{part}}_{q,n}\exp\left(2\pi in\frac{w^{\text{part}}_{ji}-\mu-w^{\text{mf}}_{\text{min}}}{x_{\text{max}}}\right)~,\nonumber\\\label{eq:WpartFourier}
\end{eqnarray}
which gives the same result as substituting $W_{s,\text{num}}^{\text{eff}}(w^{\text{eff}})$ into Eq.~\eqref{eq:PartEffectiveFieldDistribution} if $x_{\text{max}}>w^{\text{eff}}(q-1)+w^{\text{mf}}_{\text{max}}-w^{\text{mf}}_{\text{min}}$.

Having found an expression for the distribution $W^{\text{part}}_q(w^{\text{part}}_{ji})$ in terms of the distribution $W^{\text{eff}}_{s,\text{num}}(w^{\text{eff}})$ using Eqs.~\eqref{eq:WpartWeff} and~\eqref{eq:WpartFourier}, we next find a new estimate, $W^{\text{eff}}_{s+1,\text{num}}(w^{\text{eff}})$, for the distribution of $w^{\text{eff}}$, thus completing a step of the iterative solution process.
To find $W^{\text{eff}}_{s+1,\text{num}}(w^{\text{eff}})$ we use Eq.~\eqref{eq:RecursiveEquation} which we apply in several steps. 
In the first step, we calculate the distributions of $b_{ji}=F^+(\beta w^{\text{part}}_{ji})$ and $c_{ij}=F^-(\beta w^{\text{ff}}_{ij})$, then, second, we calculate the distribution of $d_{ij}=b_{ji}+c_{ij}$, and, finally, we calculate the distribution of $w^{\text{eff}}_{ij}=\beta^{-1} F^+(d_{ij})$.
The second step of this calculation is relatively straightforward, since the first $M/2$ terms of the Fourier series of the distributions of $b_{ji}$ and $c_{ij}$ lead directly to the first $M/2$ terms for the distribution of $d_{ij}$ by the convolution theorem. 
Additional techniques are required, however, for the first and last steps.
For instance, the distribution, $B(b_{ji})$, of $b_{ji}$, is written as,
\begin{eqnarray}
&&B(b_{ji})=\nonumber\\
&&=\begin{cases}\beta^{-1}W^{\text{part}}_q(\beta^{-1}F^-(b_{ji}))\left|\frac{\text{d}F^-(b_{ji})}{\text{d}b_{ji}}\right|~,&b_{ji}>0\\0~,&\text{otherwise.}\end{cases}\nonumber\\
&&=-\left(\frac{\text{d}F^-(b_{ji})}{\text{d}b_{ji}}\right)\sum_{n=-M/2+1}^{M/2}\bigg[{\tilde W}^{\text{part}}_{q,n}\nonumber\\
&&~~~~~~~~~\times\left.\exp\left(2\pi in\frac{\beta^{-1}F^-(b_{ji})-\mu-w^{\text{mf}}_{\text{min}}}{x_{\text{max}}}\right)\right]\label{eq:BfromW}
\end{eqnarray}
where we have used the property $F^-(F^+(x))=x$.
Since $b_{ji}$ depends on $w^{\text{part}}_{ji}$ in a non-linear way, the Fourier series of the distribution $B(b_{ji})$ given by Eq.~\eqref{eq:BfromW} contains an infinite number of terms and cannot be expressed exactly using our numerical approach.
We notice that, in Fourier space, the distribution $B(b_{ji})$ can be recast in the form ${\tilde B}(\omega^\prime)\Gamma[i\omega^\prime]=(2\pi)^{-1}\int {\tilde W}^{\text{part}}_q(\beta\omega)\Gamma[-i\omega]\Gamma[i(\omega+\omega^\prime)]\text{d}\omega$, where $\Gamma[z]=\int_0^\infty t^{z-1}e^{-t}\text{d}t$ is the Gamma function and the Fourier transform ${\tilde W}^{\text{part}}_q(\nu)=\int W^{\text{part}}_q(w^{\text{part}}_{ji})\exp(-i\nu w^{\text{part}}_{ji})\text{d}w^{\text{part}}_{ji}$ (${\tilde B}(\omega)$ is defined similarly in terms of $B(b_{ji})$). 
However, this expression does not lead directly to a numerically accurate method to calculate $B(b_{ji})$, due to the unwanted periodicity of the function represented by Fourier series.
Instead, we have partitioned the function $B(b_{ji})$ found from Eq.~\eqref{eq:BfromW} into $M$ bins, $B_m=\int_{\Delta^\prime (m-1/2)}^{\Delta^\prime(m+1/2)}B(b_{ji})\text{d}b_{ji}$, where $\Delta^\prime=\beta\Delta$ and $0\le m<M$.
The formula for $B_m$ in terms of ${\tilde W}^{\text{part}}_{q,n}$ is, 
\begin{eqnarray}
&&B_m
=\sum_{n=-M/2+1}^{M/2}{\tilde W}^{\text{part}}_{q,n}\nonumber\\
&&\times\int_{F^-(\Delta (m+1/2))}^{F^-(\Delta(m-1/2))}\exp\left(2\pi in\frac{w^{\text{part}}_{ji}-\mu-w^{\text{mf}}_{\text{min}}}{x_{\text{max}}}\right)\text{d}w^{\text{part}}_{ji}\nonumber\\\label{eq:NUFFT}
\end{eqnarray}
which can be evaluated for all $m$ simultaneously using a non-uniform fast Fourier transform~\cite{Dutt1993}.
The values $B_m$ are assumed to represent the function $B_m$ according to $B(m\Delta^{\prime})=B_m$.
Having found such an expression for $B(b_{ji})$, we can find the distribution of $d_{ij}$ by convolution.
A similar expression to Eq.~\eqref{eq:NUFFT} exists to link $W^{\text{eff}}_{s+1,\text{num}}(w^{\text{eff}})$ to the distribution of $d_{ij}$.
Thus, having started with a given distribution, $W^{\text{eff}}_{s,\text{num}}(w^{\text{eff}})$, in Eq.~\eqref{eq:WeffFourier}, we have found a new distribution $W^{\text{eff}}_{s+1,\text{num}}(w^{\text{eff}})$.
Repeatedly applying this process eventually leads to a fixed distribution, which is the numerical solution to Eqs.~\eqref{eq:RecursiveEquation} and~\eqref{eq:PartEffectiveFieldDistribution}.

\section{Zero- and small-disorder case}\label{sec:ExactSolution}

In this appendix, we present an analytical solution for $R(\rho)$ and the form of the phase diagram for a system with zero or small disorder.
The analysis is exact for the case of zero-disorder and approximate for small disorder.
The distribution is obtained as an approximate solution of the two coupled integral Eqs.~\eqref{eq:RecursiveEquation} and~\eqref{eq:PartEffectiveFieldDistribution}, and is based on a linear expansion of the integrand in random interaction strengths and effective fields about their mean values.

\begin{figure} 
\includegraphics[width=6cm]{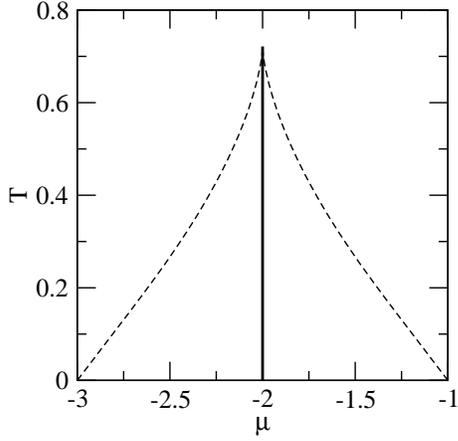}
\caption{Phase diagram for fluid condensation on a $q=4$-regular graph with homogeneous interaction strengths, $w^{\text{ff}}_0=1$, $w^{\text{mf}}_0=0$. 
The solid line represents the phase boundary and dashed lines represent the limits of metastability (spinodal lines).\label{fig:PhaseAppFig}}
\end{figure}

We start our analysis with the zero-disorder case considered in Sec.~\ref{sec:zero-disorder}.
In this case, the values of $w^{\text{eff}}_0$ and $w^{\text{part}}_0$ introduced in Sec.~\ref{sec:zero-disorder} can be found by solving the following two coupled equations,
\begin{eqnarray}
\beta w^{\text{eff}}_0&=&F^+\left[F^+\left(\beta w^{\text{part}}_0\right)+F^-\left(\beta w^{\text{ff}}_0\right)\right]\label{eq:EffZeroDisorder}\\
w^{\text{part}}_0&=&\mu+w^{\text{mf}}_0+(q-1)w^{\text{eff}}_0~,\label{eq:PartZeroDisorder}
\end{eqnarray}
which are obtained by setting all random values to be equal to their mean values in Eqs.~\eqref{eq:EffField} and~\eqref{eq:PartField}.
For temperatures below the critical value, $T_{\text{C}}$, there is a region in the parameter space of $\mu$ and $T$, within which Eqs.~\eqref{eq:EffZeroDisorder} and~\eqref{eq:PartZeroDisorder} have multiple solutions, corresponding to metastable states.
The boundaries of this region 
are at the points where Eqs.~\eqref{eq:EffZeroDisorder} and~\eqref{eq:PartZeroDisorder} have a double root, i.e. where the derivatives of the left and right hand sides of Eq.~\eqref{eq:EffZeroDisorder} with respect to $w^{\text{eff}}_0$ are equal, i.e., 
\begin{eqnarray}
&&\frac{\text{d}}{\text{d} w^{\text{eff}}_0}\beta w^{\text{eff}}_0=\nonumber\\&&\frac{\text{d}}{\text{d} w^{\text{eff}}_0}\left\{F^+\left[F^+\left(\beta w^{\text{part}}_0\right)+F^-\left(\beta w^{\text{ff}}_0\right)\right]\right\}~.\label{eq:ExactSolutionDeriv}
\end{eqnarray}
The simultaneous solution of Eqs.~\eqref{eq:EffZeroDisorder}, \eqref{eq:PartZeroDisorder} and~\eqref{eq:ExactSolutionDeriv} is given by,
\begin{eqnarray}
&&\mu^{\pm}(T)=-\frac{1}{2}qw^{\text{ff}}_0-w^{\text{mf}}_0\nonumber\\
&&\pm\Bigg\{\beta^{-1}\ln\left[
\frac{\sinh\left(\frac{1}{4}\beta w^{\text{ff}}_0-\frac{1}{2}\beta w^{\text{spin}}\right)}
{\sinh\left(\frac{1}{4}\beta w^{\text{ff}}_0+\frac{1}{2}\beta w^{\text{spin}}\right)}
\right]+(q-1)w^{\text{spin}}\Bigg\}~,\nonumber\\\label{eq:ExactSpinodal}
\end{eqnarray}
where,
\begin{eqnarray}
&&w^{\text{spin}}=\beta^{-1}\ln\left\{\frac{q}{q-2}\right.\nonumber\\
&&+\left.\frac{q-2}{2(q-1)}\left[v
+\sqrt{v\left(e^{\beta w^{\text{ff}}_0}-1\right)}\right]\right\}-\frac{1}{2} w^{\text{ff}}_0~,\label{eq:wspin}
\end{eqnarray}
and $v=\exp\left(\beta w^{\text{ff}}_0\right)-[q/(q-2)]^2$.
In Eq.~\eqref{eq:ExactSpinodal}, the two solutions $\mu^\pm(T)$ represent the upper and lower boundaries of the spinodal region at a given value of $T$ (see the dashed lines in Fig.~\ref{fig:PhaseAppFig}).
When the temperature is such that $\exp(\beta w^{\text{ff}}_0)=[q/(q-2)]^2$, the value of $w^{\text{spin}}$ given by Eq.~\eqref{eq:wspin} is zero, and the upper and lower boundaries of the spinodal coincide, $\mu^+(T)=\mu^-(T)$, meaning that this temperature corresponds to criticality. 
This condition leads to the expression for the critical temperature, $T_\text{C}$, given by Eq.~\eqref{eq:ExactSpinodal}.
For $T>T_{\text{C}}$, Eq.~\eqref{eq:ExactSpinodal} does not have any real solutions, and there are no phase transitions or coexistence for this range of temperatures.

In order to obtain an approximation of $R(\rho)$ for small normally distributed values of $w^{\text{mf}}_{i}$ and $w^{\text{ff}}_{ij}$, we expand Eqs.~\eqref{eq:EffField} and~\eqref{eq:PartField} to first order in all of the field parameters, i.e. 
\begin{eqnarray}
\beta (w^{\text{eff}}_{ij}- w^{\text{eff}}_0)&=&a[b\beta (w^{\text{ff}}_{ij}-w^{\text{ff}}_0)\nonumber\\&&+c\beta (w^{\text{part}}_{ji}-w^{\text{part}}_0)]~,\label{eq:LinearApproximationEff}\\
w^{\text{part}}_{ji}-w^{\text{part}}_0&=&w^{\text{mf}}_j-w^{\text{mf}}_0+\sum_{k/j,k\neq i}(w^{\text{eff}}_{jk}-w^{\text{eff}}_0)~,\nonumber\\
\label{eq:LinearApproximationPart}
\end{eqnarray}
where,
\begin{eqnarray}
a&=&\left.\frac{\partial F^+(x)}{\partial x}\right|_{x=F^-(w^{\text{ff}}_0)+F^+(\beta w^{\text{part}}_0)}\nonumber\\
&&=-\exp\{-[\beta w^{\text{ff}}_0+F^-(\beta w^{\text{ff}}_0)]\}~,\nonumber\\
b&=&\frac{\partial F^-(\beta w^{\text{ff}}_0)}{\partial w^{\text{ff}}_0}=[e^{-\beta w^{\text{ff}}_0}-1]^{-1}~,\nonumber\\
c&=&\frac{\partial F^+(\beta w^{\text{part}}_0)}{\partial w^{\text{part}}_0}=-[e^{-\beta w^{\text{part}}_0}+1]^{-1}~.
\end{eqnarray}
Assuming that the effective fields $w^{\text{eff}}_{ij}$ are identically normally distributed, $w^{\text{eff}}_{ij}\sim N(w^{\text{eff}}_0,\Delta_{\text{eff}}^2)$, with mean ${\overline w}^{\text{eff}}_{ij}$ and variance $\Delta_{\text{eff}}^2$, Eqs.~\eqref{eq:LinearApproximationEff} and~\eqref{eq:LinearApproximationPart} can be solved and both the mean and variance of $w^{\text{eff}}_{ij}$ found. 
The mean value does not change from the zero-disorder case, ${\overline w}^{\text{eff}}_{ij}=w^{\text{eff}}_0$, 
meaning that the solution with small disorder has the same phase diagram as the zero-disorder solution.
The variance is given by, 
\begin{equation}
\Delta_{\text{eff}}^2=\frac{a^2((b\Delta_{\text{ff}})^2+(c\Delta_{\text{mf}})^2)}{1-(q-1)(ac\Delta_{\text{eff}})^2}~.
\end{equation}

Similarly, to the analysis for the distribution of effective fields, the distribution $R(\rho_i)$ of mean cell densities can be found for small disorder by linearly expanding the argument of the $\delta$-function in Eq.~\eqref{eq:rhoDistribution} with respect to the random variables, i.e. 
\begin{equation}
\rho_i-\rho_0=\beta\rho_0(1-\rho_0)\left[w^{\text{mf}}_i-w^{\text{mf}}_0+q(w^{\text{eff}}_{ij}-w^{\text{eff}}_0)\right]~,\label{eq:LinearApproximationRho}
\end{equation}
where $\rho_0=\left\{1-\exp\left[-\beta(\mu+w^{\text{mf}}_0+qw^{\text{eff}}_0)\right]\right\}^{-1}$ is the density in the absence of disorder.
The values of $\rho_i$ are therefore normally distributed with a mean $\rho_0$ and a standard deviation,
\begin{equation}
\Delta_{\rho}=\beta\rho_0(1-\rho_0)\sqrt{\frac{qa^2\left[(b\Delta_{\text{ff}})^2+(c\Delta_{\text{mf}})^2\right]}{\left[1-(ac)^2(q-1)\right]}+{\Delta_{\text{mf}}}^2}~.\label{eq:RhoDistributionApproxWidth}
\end{equation}

We have demonstrated that for the low disorder case, the distribution of densities can be approximated by a normal distribution.
The quality of this approximation is good even for $\Delta_{\text{mf}}/w^{\text{ff}}_0=0.6$ as can be seen by comparing the data points according to the approximate normal solution (circles in Fig.~\ref{fig:DistributionsInterDis}(a)) with the results of the exact numerical solution (lines) and of Metropolis dynamics simulations (crosses). 

\section{Method of Monte Carlo simulations}\label{sec:MetropolisAppendix}
 
The dynamical simulations of the evolution of a system of cells is performed using Metropolis dynamics~\cite{Metropolis1953}.
As explained in Sec.~\ref{sec:Metropolis}, in this dynamics, a random cell, $i$, is selected out of $N$ cells with probability $P_{\text{s}}=1/N$ and its state is changed with the probability given by $P_i=\min[1,\exp(-\beta \Delta{\cal H})]$.
If the cell does not change state, the transition is referred to as a null transition.
At low temperatures, starting from any initial state, the system rapidly moves into a metastable state.
After this, most transitions are null transitions and nothing changes in the system for long periods of time within the standard Metropolis scheme.
To improve the computational performance of our simulations, we use a level 1 Monte Carlo Adsorbing Markov Chain (MCAMC) method~\cite{LandauBinderBOOK,Novotny1995a,Novotny1995b}, which effectively skips the null transitions. 

In our implementation of the MCAMC, we calculate the probability, $P_i$, that a spin $i$ will change state if it is chosen.
By calculating the value of $\sum_{i=1}^NP_{\text{s}}P_i$ we can find the probability that the next transition is a null transition, 
\begin{equation}
P_{\text{null}}=1-\sum_{i=1}^NP_{\text{s}}P_i~,
\end{equation}
If the next transition is a null transition and no parameters of the system change between different transitions, then the transition following the next one will also be a null transition with probability $P_{\text{null}}$, and so on.
Thus, we can find that the probability, $P_n$, that sequence of $n-1$ null transitions is followed by a non-null transition, i.e.
\begin{equation}
P_n=P_{\text{null}}^{n-1}(1-P_{\text{null}})~.\label{eq:NullSkips}
\end{equation}
In the Metropolis simulation, we first choose to skip a random number of transitions, $n$, drawn from the distribution Eq.~\eqref{eq:NullSkips} and then change the state of a cell $i$, randomly chosen
according to the probability distribution,
\begin{equation}
P({i}|{\text{non-null}})=\frac{P_i}{\sum_{j=1}^NP_j}.
\end{equation}
Finally, the probabilities $P_i$ for the cell which changed state and all of its neighbours are updated to reflect their new environment.
Numbers can be selected from such a distribution in $O(\log n)$ computational steps by using a binary tree (see e.g. Ref.~\cite{Fishman1993}).

Within such an implementation of the MCAMC method, and for the case of Metropolis dynamics, a changing chemical potential can be modelled without having to recalculate every value of $P_i$.
To do this, we do not store the values of $P_i$ explicitly, but instead store the values, 
\begin{equation}
P_i^{(0)}=
\begin{cases}
\exp(-\beta (2\tau_i-1)f^{(0)}_i)&\text{if}~(2\tau_i-1)f_i\ge 0\\
1&\text{if}~(2\tau_i-1)f_i<0~.
\end{cases}
\end{equation}
with the local field, $f_i$, and reference field, $f_i^{(0)}$, given by,
\begin{equation}
f_i=\mu+\sum_{j/i}\tau_jw^{\text{ff}}_{ij}+w^{\text{mf}}_i~,\quad f_i^{(0)}=f_i-(\mu-\mu_0)~.
\end{equation}
The values of $P_i$ are calculated according to the following equation,
\begin{equation}
P_i=P_i^{(0)}U_i~,\label{eq:ScalePI} 
\end{equation}
where the values of $U_i$ are given by,
\begin{equation}
U_i=\left\{
\begin{array}{ccc}
a&\text{if}~\tau_i=0,f_i\le 0&(a)\\
a^{-1}&\text{if}~\tau_i=1,f_i\ge 0&(b)\\
1&\text{otherwise,}&(c)
\end{array}\right.\label{eq:NFoldCases}
\end{equation}
and the scaling factor $a=\exp[\beta(\mu-\mu_0)]$.
The value of $\mu_0$ is arbitrary, but it is chosen to be such that $a$ is within the range of the floating point representation. 
According to Eq.~\eqref{eq:NFoldCases}, 
the different cells can be separated into three distinct classes, depending on their state $\tau_i$ and the local field $f_i$.
These classes correspond to cells which are stable and unoccupied (${\cal C}_{\text{su}}$, see Eq.~\eqref{eq:NFoldCases}(a)), stable and occupied (${\cal C}_{\text{so}}$, see Eq.~\eqref{eq:NFoldCases}(b)) and unstable (${\cal C}_{\text{u}}$, see Eq.~\eqref{eq:NFoldCases}(c)).

Separate binary trees are used to generate random variates, $i$, from the distributions, $P_i^{(0)}$, corresponding to cells in each of the three classes. 
The sums, $P^{\text{so}}=\sum_{i\in{\cal C}_{\text{so}}}P^{(0)}_i$, $P^{\text{su}}=\sum_{i\in{\cal C}_{\text{su}}}P^{(0)}_i$ and $P^{\text{u}}=\sum_{i\in{\cal C}_{\text{u}}}P^{(0)}_i$ are also stored separately, being the roots of these trees. 
The algorithm for simulating Metropolis dynamics is then as follows.
First, the probability of the next transition being a null transition is calculated using $P_{\text{null}}=aP^{\text{su}}+a^{-1}P^{\text{so}}+P^{\text{u}}$.
The next $n$ transitions are then skipped, where $n$ is distributed according to $P_n$ in Eq.~\eqref{eq:NullSkips}
Second, the class of cell which changes state on the next transition is chosen according to the probabilities,
\begin{eqnarray}
P(\text{so}|\text{non-null})&=&\frac{a^{-1}P^{\text{so}}}{aP^{\text{su}}+a^{-1}P^{\text{so}}+P^{\text{u}}}\nonumber\\
P(\text{su}|\text{non-null})&=&\frac{aP^{\text{so}}}{aP^{\text{su}}+a^{-1}P^{\text{so}}+P^{\text{u}}}\nonumber\\
P(\text{u}|\text{non-null})&=&\frac{P^{\text{u}}}{aP^{\text{su}}+a^{-1}P^{\text{so}}+P^{\text{u}}}~.
\end{eqnarray}
The particular cell from the chosen class which changes state is selected by using the binary tree method.
The final step is to update the value of $P_i^{(0)}$ and the class for the cell which changed state and for all of its neighbours.

An additional tree should be introduced in order to account for any cells which change class as $\mu$ changes.
For concreteness, consider the case when $\mu$ increases.
In this case, some cells with a negative local field may begin to experience a positive local field and thus their class should be changed (see Eq.~\eqref{eq:NFoldCases}.
In order to determine whether any cell changes between the classes, we maintain a list of $f_i^{(0)}$ for all of the cells which have negative local fields $f_i$. 
After increasing $\mu$, we find the largest value of $f_i^{(0)}$ in the list and check whether the corresponding local field, $f_i=f_i^{(0)}+(\mu-\mu_0)$, is positive or negative.
If it is positive, the value of $f_i^{(0)}$ for that cell is removed from the list of negative fields and the class together with the values of $P_i^{(0)}$ are corrected for the cell.
Such a procedure is repeated until the highest local field in the list of negative local fields is negative.
The list of negative local fields is also stored in a binary search tree meaning that this operation takes $O(\log N)$ computational steps.

Within this more complex version of the MCAMC method, a change in $\mu$ can be performed by changing the value of $a$ and checking the list of negative local fields. 
The technique described above allows the value of $\mu$ to be smoothly varied with time and performs better than the simple Metropolis method for low temperatures $T\lesssim 0.5$, when more than approximately $9/10$ of transitions are null transitions. 
The main memory requirement for this method is for the four binary trees used. 
Each binary tree has two nodes for each site, so the memory requirement scales linearly with system size.
Note that this method of accelerating the simulations can only be achieved for Metropolis dynamics and not Glauber dynamics because Eqs.~\eqref{eq:ScalePI} and~\eqref{eq:NFoldCases} hold only for Metropolis dynamics.
Indeed, for Metropolis dynamics, a change of $\mu$ results in a change of $P_i$ by $U_i$ where $U_i$ can only take three values, $a$, $a^{-1}$ and $1$.
Conversely for Glauber dynamics, the value of $U_i=[\cosh(\beta f_i^{(0)}s_i)/\cosh(\beta f_is_i)]\exp[-\beta (\mu-\mu_0)s_i]$, which is, in general, different for every cell and thus requires recalculation of every value of $P_i$ for any new value of $\mu$. 


%

\end{document}